# Advanced materials for magnetic cooling: fundamentals and practical aspects

M. Balli[1, 2]*, S. Jandl[1, 2], P. Fournier[1, 2, 3], A. Kedous-Lebouc[4]

[1] Institut Quantique, Université de Sherbrooke, J1K 2R1, QC, Canada.

[2] Regroupement Québécois sur les Matériaux de Pointe, Département de Physique, Université de Sherbrooke, J1K 2R1, QC, Canada.

[3] Canadian Institute For Advanced Research, Ontario M5G 1Z8, Canada.

[4] G2Elab, Grenoble Institute of Technology, 21 avenue des martyrs, 38031 Grenoble CEDEX1, France.

*Mohamed.balli@uir.ac.ma

**ABSTRACT.** Over the last two decades, the research activities on magnetocalorics have been exponentially increased leading to the discovery of a wide category of materials including intermetallics and oxides. Even though the reported materials were found to show excellent magnetocaloric properties on laboratory scale, only a restricted family among them could be upscaled toward industrial levels and implemented as refrigerants in magnetic cooling devices. On the other hand, in the most of reported reviews, the magnetocaloric materials are usually discussed in terms of their adiabatic temperature and entropy changes ($\Delta T_{ad}$ and $\Delta S$), which is not enough to get more insight about their large scale applicability. In this review, not only the fundamental properties of recently reported magnetocaloric materials are discussed but also their thermodynamic performance in functional devices. The reviewed families particularly include $Gd_{1-x}R_x$ alloys, $LaFe_{13-x}Si_x$, $MnFeP_{1-x}As_x$ and $R_{1-x}A_xMnO_3$ (R = lanthanide, A = divalent alkaline earth) –based compounds. Other relevant practical aspects such as mechanical stability, synthesis and corrosion issues are discussed. In addition, the intrinsic and extrinsic parameters that play a crucial role in the control of magnetic and magnetocaloric properties are regarded. In order to reproduce the needed magnetocaloric parameters, some practical models are proposed. Finally, the concepts of the rotating magnetocaloric effect and multilayered magnetocalorics are introduced.

**TABLE OF CONTENTS:**

# I. INTRODUCTION

With the growing concerns about global warming, negative impact of synthetic refrigerants on the environment and energy resources scarcity, the major challenge of the refrigeration industry is the reduction of energy consumption and harmful gas emissions. In fact, the refrigeration plays an increasingly vital role in many domains of our everyday life such as food preservation and production, air-conditioning, gas liquefaction, preservation of human organs and much more. Until the year 2008, there are about 1 billion domestic cooling systems in use worldwide and this is constantly expanding [1]. For example, between 1996 and 2008 (over 12 years), the number of household refrigerators has increased by approximately 100 % [1]. According to S. Pearson [2], about 15 % of the world's electricity consumption is used in refrigeration and air-conditioning systems, while in developed countries this percentage reaches about 30 % and expected to markedly increase if there are some deficiencies in the cooling devices [3, 4]. On the other hand, within the conventional refrigeration, the cooling process is performed by employing a vapour-compression cycle of some harmful fluids such as chlorofluorocarbons (CFCs), hydrochlorofluorocarbons (HCFCs) and hydrofluorocarbons (HFCs). As CFCs and HCFCs refrigerants were found to be mainly responsible for ozone layer depletion [5], the Montreal protocol was universally adopted for the purpose of restricting their utilization [6]. In response to the regulation of ozone depleting substances, the production and use of HFCs have significantly increased as substitutes for CFCs refrigerants [7]. However, HCFs belongs to a family of greenhouse gases (GHG) with an effective global warming potential (GWP) that is thousands of times greater than that of $CO_2$ [8]. With regards to GWP, in 1997, a global treaty to reduce emissions of greenhouse gases was adopted in Kyoto (Kyoto protocol) [9]. In recent years, many countries around the world, including the European Union (EU), Japan, USA and China begun to unveil new rules to phase out GWP gases. On the other hand, worldwide research and developments have been stimulated to deal with the drawbacks of traditional cooling methods.

Based on the magnetocaloric effect (MCE), magnetocaloric refrigeration is currently considered as a promising substitution for standard cooling techniques since it enables to completely eliminate fluorinated gases (F-gases) while presenting high energy efficiency [10]. Over the last two decades, the research activities on both magnetocaloric materials and magnetic cooling devices have been exponentially increased. In addition, the creation of an international conference (Thermag) completely dedicated to the magnetic refrigeration with the purpose of consolidating the collaboration between scientists studying magnetocaloric materials and devices designers, unveils the promising future of this "green" technology.

The magnetocaloric effect which provides the basis of the magnetic cooling is a well-known phenomenon and has been widely implemented in the past to reach very low temperatures. Nearly a century ago, changes in the nickel temperature when varying the external magnetic field were originally discovered by Pierre Weiss and Auguste Piccard in 1917, during their study of magnetization as function of temperature and magnetic field near the magnetic phase transition [11]. The observed temperature increase was then called by Weiss and Piccard *le phénomène magnétocalorique* (magnetocaloric phenomenon). However, it is worth noting that Langevin has already demonstrated in 1905 the possibility of paramagnetic substances to release heat during a reversible modification of their magnetization [12]. In the late of 1920s, a major advance occurred when Debye (1926) [13] and Giaugue (1927) [14] independently proposed an additional thermodynamic explanation of the magnetocaloric effect and suggested the refrigeration process to obtain low temperatures by using adiabatic demagnetization of paramagnetic salts. The concept was experimentally implemented for the first time a few years later when in 1933 Giaugue and MacDougall [15] attained a temperature of 0.25 K by demagnetizing adiabatically the gadolinium sulfate, $Gd_2(SO_4)_8H_2O$, at the temperatures of liquid helium. A solenoid producing a field of about 0.8 T and 61 g of $Gd_2(SO_4)_8H_2O$ were used in the experimental device. This major work led to a Nobel Prize awarded to Giauque and MacDougal in 1949. Between 1933 and the beginning of 1970s, most of published studies were devoted to low temperature (below 20 K) cooling [16]. However, the great step towards the magnetic cooling at room-temperature was bridged in 1976 when Brown [17] demonstrated the possibility to utilize the magnetocaloric effect of gadolinium (Gd) to produce a significant cooling effect around 294 K. In Brown's magnetic cooling system, one mole of 1mm-thick Gd plates, separated by screen wires was arranged in a cylindrical assembly. A fluid constituted of 80 % water and 20 % ethyl alcohol was used for the heat exchange. The thermal effect was generated by an alternating 7 T field produced by a water-cooled electromagnet. After about 50 magnetic Stirling-cycles, a temperature span of 47 K was obtained between the hot end (46 °C) and the cold end (-1 °C). By using the same magnetic refrigerator [18], Brown et al reached a temperature span of 80 K, between 248 K (-26 °C) and 328 K (54 °C). For this purpose, 0.9 kg of Gd formed in 1mm thick plates and a heat transfer fluid constituted of 50 % ethanol and 50 % water were utilized. Following the pioneer works of Brown, several works were conducted with the aim to render the magnetic refrigeration technology more attractive in the near room-temperature range [16, 19, 20, 21].

In the late of 1990s, two major works which generated a huge of interest in the field, occurred when Ames Laboratory [22] and Astronautic Corporation of America [10] unveiled a new performant magnetocaloric material for room temperature tasks and a competitive magnetic cooling device, respectively. In 1998, Zimm et al [10] reported a successful operating device, demonstrating that magnetocaloric cooling is a competitive technology for both domestic and industrial uses. Using a bed of gadolinium spherical particles as refrigerants and a field of 5 T produced by a superconducting magnet, the authors were able to achieve a maximum temperature span of 38 K and cooling powers exceeding 500 watts at coefficients of performance (COP) larger than 6. They also showed that 60 % of Carnot efficiency can be attained with 281 K to 291 K temperature span. In 1997, Pecharsky and Gschneidner [22] reported the so-called giant magnetocaloric effect (GMCE) in $Gd_5Si_2Ge_2$-based compounds around ambient temperature. The observed GMCE was the result of the first order magneto-structural transformation associated with the transition from the ferromagnetic to the paramagnetic phase, occurring close to 273 K. The obtained maximum entropy change is about twice as large as that of Gd considered as a reference for magnetocaloric materials. This discovery has remarkably stimulated both fundamental and applied researches increasing exponentially the number of works in the field [16].

It is worth noting that a giant MCE was reported by Annaorazov et al [23] in $Fe_{0.49}Rh_{0.51}$ about 5 years before the Pecharsky and Gschneidner work [22]. The investigated compound undergoes a field-induced antiferromagnetic-ferromagnetic first order magnetic phase transition at~313 K. The application of a magnetic field of about 2 T to a sample of $Fe_{0.49}Rh_{0.51}$ causes a large temperature change of 13 K. Until now, the $Fe_{0.49}Rh_{0.51}$ compound can be considered as the best magnetocaloric material in terms of the adiabatic temperature change ($\Delta T_{ad}$). The little practical interest given to $Fe_{0.49}Rh_{0.51}$ based alloys can be mainly attributed to the scarcity of Rh (excessively expensive) and the irreversibility of the magnetocaloric effect with regard to the magnetization-demagnetization process. However, Manekar and Roy [24] have demonstrated that the reproducibility of MCE in Fe-Rh alloys is possible if the magnetic field-temperature history of the sample is taken into account by using the second isothermal magnetization cycle (envelope) to calculate ΔS rather than the virgin magnetization curves. This approach was utilized by Barua et al [25] to evaluate the isothermal entropy change in FeRh-based ternary compounds.

After the discovery of the GMCE in $Gd_5Si_2Ge_2$ [22], intensive studies were devoted to the development of "useful or practical" magnetocaloric materials and understanding the physics behind their properties. Since then, a wide variety of advanced magnetocalorics with a GMCE such as La(Fe, Mn, Co, Mn)$_{13-x}$Si$_x$(H,N, C)$_y$ [26-40], MnAs$_{1-x}$Sb$_x$ [41], Fe$_2$P-types compounds (MnFeP$_{1-x}$As$_x$) [42, 43], Ni-Mn-based Heusler [44, 45] and La$_{1-x}$Ca$_x$MnO$_3$ manganites [46] was reported in the literature. Following, a parallel effort was paid to design new types of efficient magnetic refrigerators, giving rise to pre-industrial systems [47-50]. However, the gap to be bridged in going from laboratory samples to a competitive device that meets the market needs is demanding. In fact, the magnetocaloric material must answer a series of requirements before its direct implementation such as, sufficiently large MCE on a wide temperature range, high thermal conductivity, low specific heat, low hysteresis effect, high electrical resistance, high resistance against oxidation and corrosion, mechanical stability and safe constituent elements. Thus it is very difficult to find a material that combines all these characteristics. On the other hand, before entering the market, magnetic cooling refrigerators must also satisfy a number of requirements such as household standards, reasonable price and size and attractive design [51].

An example of magnetic refrigerator[8, 49] is reported in Figure 1. As shown, thermal effects can be generated by moving the magnetocaloric material (MCM) inside and outside of a magnet via a linear actuator. The heat exchange is usually achieved by a moving carrier fluid such as water.

In this review destined to scientists, engineers, undergraduate and graduate students, different aspects of the magnetocaloric effect are explained. Recent progresses in relation with the implementation of relevant advanced magnetocaloric materials in magnetic refrigerators are reviewed. Some practical aspects of magnetocaloric materials such as stability issues are also considered.

## II. FUNDAMENTALS

### A. Magnetocaloric effect: physical origin

As outlined in section I, the magnetocaloric effect exhibited by certain magnetic substances is the basis of the magnetic cooling. It can be defined as the thermal response (heating or cooling) of a magnetic material under the effect of an external magnetic field. However, caloric effects could be also obtained in solid state materials by manipulating their degrees of freedom such as electric polarization, strain and volume through a variable external field [20]. In the absence of any physical coupling phenomenon, the corresponding fields to electric polarization, volume and strain are electric field, pressure and stress, respectively. Their changes lead to electrocaloric (ECE), barocaloric (BCE) and elastocaloric (ElCE) effects, respectively. For MCE, the induced temperature change is the result of magnetothermal interplay between the magnetic moments and the atomic lattice (phonons). At constant pressure, the full entropy of a magnetic substance is a function of both magnetic field (H) and temperature [52, 53]. It consists of magnetic ($S_m$), lattice ($S_{Lat}$), and electronic ($S_{El}$) contributions and can be expressed as follow,

$$S(T,H) = S_{Lat}(T,H) + S_{El}(T,H) + S_m(T,H) \qquad (1)$$

In general, the magnetic field dependence of $S_{Lat}$ and $S_{El}$ is neglected, while $S_m$ is very sensitive to the external magnetic field. On the other hand, the contribution from electrons to the magnetocaloric effect is usually neglected in systems that show a localized magnetism such as rare earth-based materials. When the magnetic field is isothermally applied, the magnetic moments arrangement is reorganized which consequently enhances or reduces the magnetic entropy part, depending on the materials initial magnetic state. For typical ferromagnets and paramagnets, the application of magnetic field (increasing field from $H_I$ to $H_F$) tends to orient the magnetic moments along the field direction (see Fig.2), making the magnetic material more ordered. This decreases the magnetic entropy and consequently the full entropy by:

$$\Delta S(T, H_I \rightarrow H_F) = S_F(T, H_F) - S_I(T, H_I) \qquad (2)$$

with $H_F > H_I$ and $S_F < S_I$. In adiabatic conditions, the full entropy is conserved, i.e. $S_F(T_F, H_F) = S_I(T_I, H_I)$. Consequently, the magnetic entropy loss is compensated by the change of the quantity $S_{Lat} + S_{El}$ in the opposite way, increasing then the material temperature (Fig.2) by:

$$\Delta T_{ad}(T_I, H_I \rightarrow H_F) = T_F(S_F) - T_I(S_I) \qquad (3)$$

with $S_F = S_I$.

In a reversible process, the magnetic moments return to their random state when removing the applied magnetic field. In this case, the magnetic entropy increases and the material is forced to cool down. In Fig.2 is schematically plotted the resulting MCE of a typical ferromagnetic material (gadolinium) in terms of $\Delta T_{ad}$ and $\Delta S$ for an initial temperature equal to its Curie point (294 K) and a magnetic field changing from 0 to 5 T.

The quantities $\Delta T_{ad}$ and $\Delta S$ are amongst the most used figures of merit to identify the potential of magnetocaloric materials. The intrinsic and extrinsic parameters that importantly affect their behaviors are discussed in sections II-B to II-D. On the other hand, it is worth noting that a negative change of temperature can be exhibited by certain materials when a magnetic field is applied, which contrast with that of ordinary ferromagnetic systems [23, 54-56]. This is called the negative (or inverse) magnetocaloric effect and mainly concerns the antiferromagnets (AF). The latter cool down when magnetized and heat up when demagnetized. This is because the application of an external magnetic field changes their magnetic state from an ordered phase (AF with lower energy level) to a less-ordered phase (Ferro or Para for example), increasing the material magnetic entropy. In adiabatic conditions, the material's temperature decreases to compensate for this variation. In the absence of an external magnetic field, the magnetic lattice returns to its ordered state, increasing the system temperature, according to equation 1.

It is worth noting that the discovery of the magnetocaloric effect was widely attributed to the German physicist Emil Warburg. The Warburg's paper published in 1881 [57], is systematically cited in the most majority of works in relation with the magnetic refrigeration. However, according to a recently reported work by Anders Smith [58] entitled "Who discovered the magnetocaloric effect", it clearly seems that the first experimental measurement of the MCE was done by Weiss and Piccard in 1917 [11]. In fact, Warburg neither measured the MCE in terms of temperature or heat. In his famous work [57], the magnetization of iron wire is measured in increasing and decreasing magnetic field around the room-temperature, which is equivalent to a hysteresis cycle. Accordingly, he stated that the magnetic irreversibility results in heat dissipation in the ferromagnetic body [58]. It should be noted that Thomson (in 1860) was the first to demonstrate the physics behind the magnetocaloric effect [58, 59]. Based on thermodynamics considerations, he predicted that iron will heat up if magnetized and cool down when demagnetized. Besides, the thermodynamic origin of the MCE in paramagnets was also discussed by Langevin [12], almost 45 years after Thomson work. A detailed work tracing the history of the MCE can be found in A. Smith [58].

B. **Thermodynamic aspects**

In order to well understand the magnetocaloric effect behaviour, it is useful to recall the thermodynamic properties of a magnetic material plunged in a magnetic field $H$ at a temperature $T$ and under a pressure $P$ [52, 53, 60]. The critical thermodynamic behaviour of a magnetic system can be investigated in the framework of Gibbs free energy G. This latter can be expressed as follow:

$$G = U - TS + PV - MB \text{ (with } B = \mu_0 H \text{)} \tag{4}$$

where U is the internal energy, S is the full entropy, V is the volume and M is the magnetization. Its total differential is given by:

$$dG = dU - TdS - BdM + PdV - SdT - MdB + VdP \tag{5}$$

Since the free energy G is a state function, its total differential has the following form:

$$dG = \left(\frac{\partial G}{\partial T}\right)_{P,B} dT + \left(\frac{\partial G}{\partial B}\right)_{P,T} dB + \left(\frac{\partial G}{\partial P}\right)_{B,T} dP \tag{6}$$

The generalized thermodynamic forces V, S and M can be then identified by the following equations:

$$V = \left(\frac{\partial G}{\partial P}\right)_{B,T}, \quad S = -\left(\frac{\partial G}{\partial T}\right)_{P,B}, \quad M = -\left(\frac{\partial G}{\partial B}\right)_{P,T} \tag{7}$$

where T, B and P are taken as the external variables. Based on equation 7, we obtain the following relation:

$$\left(\frac{\partial M}{\partial T}\right)_{P,B} = -\frac{\partial}{\partial T}\left(\left(\frac{\partial G}{\partial B}\right)_{P,T}\right)_{P,B} = -\frac{\partial}{\partial B}\left(\left(\frac{\partial G}{\partial T}\right)_{P,B}\right)_{P,T} = \left(\frac{\partial S}{\partial B}\right)_{P,T} \tag{8}$$

Then the thermodynamic Maxwell relation that links the entropy change to the bulk magnetization, the magnetic field and the temperature is obtained. Under a magnetic field changing from 0 to H (B = μ₀H), the isothermal entropy change can be written as the integral form of the Maxwell relation:

$$\Delta S(T, 0 \to B) = \int_0^B \left(\frac{\partial M}{\partial T}\right)_{P,B'} dB' \tag{9}$$

This equation shows that the isothermal entropy change is proportional not only to the magnitude of the magnetic field but depends strongly on the nature of the magnetic phase transition. In the case of materials exhibiting a first order character of the magnetic phase transition, i.e. a rapid variation of the order parameter as a function of the temperature (discontinuous change), the derivative of magnetization with respect to the temperature becomes larger, leading to large values of ΔS. Usually, the obtained ΔS is peaked on a narrow working temperature range. In contrast, for second order transition materials, ΔS reveals less marked feature values but on a wide magnetocaloric working temperature range. However, the isothermal entropy change can be also determined from specific heat measurement by using the second law of thermodynamics:

$$\left(\frac{\partial S}{\partial T}\right)_{B,P} = \frac{C_P(T,B)}{T} \tag{10}$$

where $C_P$ is the total specific heat. The integration yields to

$$S(T,B) = S_0 + \int_0^T \frac{C_P(T',B)}{T'} dT' \tag{11}$$

At absolute zero, the full entropy $S_0$ is usually considered to be 0. In this case, the isothermal entropy change corresponding to the field variation from 0 to B, can be expressed as follow:

$$\Delta S(T, 0 \rightarrow B) = \int_0^T \frac{C_P(T',B) - C_P(T',0)}{T'} dT' \tag{12}$$

Besides, the infinitesimal entropy change $dS$ for an isobaric process is given by:

$$dS = \left(\frac{\partial S}{\partial T}\right)_B dT + \left(\frac{\partial S}{\partial B}\right)_T dB \tag{13}$$

By using the thermodynamic Maxwell relation (Eq.8) and the second law of thermodynamics (Eq.10), equation 13 becomes:

$$dS = \frac{C_B}{T} dT + \left(\frac{\partial M}{\partial T}\right)_B dB \tag{14}$$

In a reversible adiabatic process (dS = 0), the integration of the above equation yields to the second parameter that measures the magnetocaloric effect, namely the adiabatic temperature change $\Delta T_{ad}$,

$$\Delta T_{ad}(T, 0 \to B) = -\int_0^B \frac{T}{C_P(T,B)} \left(\frac{\partial M}{\partial T}\right)_{B'} dB' \tag{15}$$

According to equation (15), the adiabatic temperature change is inversely proportional to the specific heat. The lower the specific heat is the higher $\Delta T_{ad}$ may be. However, the sign and the nature of the magnetocaloric effect i.e. negative (inverse) or conventional, are governed by the sign of the derivative of magnetization with respect to temperature (dM/dT). For ferromagnets and paramagnets, the magnetization decreases with increasing temperature (dM/dT < 0) which results in a conventional MCE ($\Delta T_{ad}$ > 0). For magnetocaloric materials presenting AF-F or AF-Para phase transitions, the magnetization increases with temperature (dM/dT > 0) and hence, the MCE is negative ($\Delta T_{ad}$ < 0).

According to the second law of thermodynamics and Maxwell relation, the equation below can be obtained.

$$\frac{\partial}{\partial B}\left(\frac{C_P}{T}\right) = \frac{\partial}{\partial B}\left(\frac{\partial S}{\partial T}\right) = \frac{\partial}{\partial T}\left(\frac{\partial S}{\partial B}\right) = \frac{\partial}{\partial T}\left(\frac{\partial M}{\partial T}\right) \tag{16}$$

In the case of materials showing a second order magnetic transition, dM/dT shows usually a non-peaked maximum (or minimum) in the magnetic phase transformation zone and hence, $\frac{\partial}{\partial B}\left(\frac{C_P}{T}\right) \cong 0$ because $\frac{\partial}{\partial T}\left(\frac{\partial M}{\partial T}\right) = 0$. This means that the term $\frac{T}{C_P}$ is magnetic field independent. Consequently, the adiabatic temperature change can be approached by

$$\Delta T_{ad} = -\frac{T}{C_P} \Delta S \tag{17}$$

From equation 17, larger MCE ($\Delta T_{ad}$) can be expected for materials with high entropy change and low total specific heat.

C. **Practical models for magnetocaloric materials**

The theoretical prediction of magnetization, specific heat, adiabatic temperature change and the isothermal entropy change is very useful from both fundamental and practical points of view. In addition to understand the mechanisms behind the magnetocaloric effect, the modelling process enables to reproduce the needed parameters for the design of functional magnetic cooling devices. For example, the magnetization data allows the prediction of involved magnetic interactions (forces and torques) between the magnetic field source and the magnetocaloric regenerator. Consequently, the needed work can be well simulated. On the other hand, the reproduced $C_P$, $\Delta T_{ad}$ and $\Delta S$ are crucial parameters for the simulation of the efficient AMR cycle (active magnetic refrigeration)[52, 61], that is usually used by magnetic cooling devices. So, in this section we report some practical models essentially based on the molecular mean field theory (MFT). The proposed models can be used to quantify the magnetocaloric properties of materials that exhibit both first and second order magnetic phase transitions.

In the absence of magneto-volume effects (second order transitions), the magnetization behavior of systems presenting localized interacting magnetic moments can be well described as a function of temperature and external magnetic field by the Brillouin function [52, 62-64] given by

$$\sigma = \frac{M}{M_0} = B_J(y) = \frac{2J+1}{2J}\coth\left(\frac{2J+1}{2J}y\right) - \frac{1}{2J}\coth\left(\frac{1}{2J}y\right) \tag{18}$$

with

$$y = \frac{1}{T}\left[3T_C\left(\frac{J}{J+1)}\right)\sigma + \frac{g_J \mu_B J}{k}B\right] \tag{19}$$

$M_0$ is the saturation magnetization, $\sigma$ is the relative magnetization, J is the angular momentum quantum number, $T_C$ is the Curie temperature, $\mu_B$ is the Bohr magneton, $g_J$ is the Landé factor and k is the Boltzman constant. The first and second terms in the y function describe the exchange interactions and the Zeeman energy, respectively. For complex magnetic substances, the $g_J$ parameter is usually assumed, while J can be deduced from the saturation magnetization, $M_0 = J* g_J* \mu_B$. In Figure 3 we report an example of the calculated magnetization reported for the $La_2NiMnO_6$ double perovskite [65] as a function of temperature at 5 T. As shown, the magnetic behaviour can be well described in the framework of mean field calculations.

It is worth noting that the binary rare earth alloys $R_xR'_{1-x}$ (R, R' = magnetic rare earth) are widely used as refrigerants in magnetic cooling systems. Their implementation enables to optimize the magnetocaloric devices and to increase their thermodynamic performance. In the case of $R_xR'_{1-x}$ alloys, the parameters J, $g_J$, and $T_C$ can be obtained from the de Gennes model [66] by using the following relationships:

$$G_{R-R'} = xG_R + (1-x)G_{R'} \text{ and } \mu_{R-R'}^2 = x\mu_R^2 + (1-x)\mu_{R'}^2 \tag{20}$$

with $G = (g_J - 1)^2 J(J+1)$ is the de Gennes factor, $\mu = g_J\sqrt{J(J+1)}$ the effective magnetic moment, $x$ and $1-x$ are the concentrations of R and R' in the alloy $R_xR'_{1-x}$, respectively. The Curie temperature can be evaluated through the relation $Tc = 46G^{2/3}$. The de Gennes model was successfully applied to several alloys such as Gd-Dy [67] and Gd-Tb [68].

The temperature and magnetic field dependence of the isothermal entropy change can be calculated by using the expression for the magnetic entropy $S_m$ as reported in Smart model [52, 62-64]

$$S_m = R\left[\ln\left(\frac{\sinh(\frac{2J+1}{2J}y)}{\sinh(\frac{y}{2J})}\right) - yB_J(y)\right] \tag{21}$$

with R is the universal gas constant. Under a magnetic field variation from $B_I$ to $B_F$, the corresponding entropy change can be expressed as

$$\Delta S(T, \Delta B = B_F - B_I) = S_m(T, B_F) - S_m(T, B_I) \tag{22}$$

The adiabatic temperature change $\Delta T_{ad}$ can be determined form the full entropy (as shown in Fig.2) consisting of the magnetic entropy $S_m$, the electronic entropy $S_{el}$ and the lattice entropy $S_L$ ($S = S_m + S_{el} + S_L$). The electronic entropy is given by the standard relation [52]

$$S_{el} = a_e T \tag{23}$$

where $a_e$ is the electronic heat capacity coefficient. The lattice entropy is obtained according to the Debye model[52]

$$S_L = -3R \ln\left[1 - \exp(-\frac{T_D}{T})\right] + 12R\left(\frac{T}{T_D}\right)^3 \int_0^{\frac{T_D}{T}} \frac{x^3}{\exp(x)-1} dx \qquad (24)$$

where $T_D$ is the Debye temperature. For a magnetic field change of $\Delta B = B_F - B_I$, the induced $\Delta T_{ad}$ is then given by :

$$\Delta T_{ad}(T, \Delta B = B_F - B_I) = T_F(S_F) - T_I(S_I) \qquad (25)$$

with $S_F = S_I$. However, $\Delta T_{ad}$ can be also calculated from $\Delta S$ values and specific heat data by using the equation 17. For this purpose, the total specific heat ($C_p = C_m + C_{el} + C_L$) with magnetic, electronic and lattice contribution must be determined. The magnetic specific heat is given by

$$C_m = T \frac{\partial S_m}{\partial T} \qquad (26)$$

while the heat capacity associated with lattice vibrations is given by the Debye model [52].

$$C_L = 9R\left(\frac{T}{T_D}\right)^3 \int_0^{\frac{T_D}{T}} \frac{x^4 e^x}{(e^x - 1)^2} dx \qquad (27)$$

The electronic specific heat ($S_{el} = C_{el}$) can be deduced from the equation 23.

The equations 18 and 19 are usually used to reproduce the magnetic and magnetocaloric parameters of materials with second order transitions. However, in materials that exhibit first-order phase transitions associated with magneto-structural transformations, the magnetic exchange interactions are very sensitive to interatomic distances. In this case, the Curie temperature is volume dependent and can be expressed in the framework of the Bean-Rodbell model [69]:

$$T_C = T_0(1 + \beta(V - V_0)/V_0) \qquad (28)$$

where $T_0$ is the Curie temperature for a non-compressible lattice, $V$ is the volume and $V_0$ is the volume in the absence of exchange interactions, $\beta$ is the slope for the volume dependence of $T_C$. In this situation, the expression of the magnetization and magnetocaloric parameters can be found via the Gibbs free energy given by [69-76]

$$G = G_{exch} + G_{Zeeman} + G_{elastic} + G_{entropy} + G_{press} \tag{29}$$

where $G_{exch}$, $G_{Zeeman}$, $G_{elastic}$, $G_{entropy}$ and $G_{press}$ denote the exchange interactions, the Zeeman energy, the elastic energy, the entropy term and the pressure term. They are expressed as follow

$$G_{exch} = -\frac{3J}{2(J+1)} NkT_C \sigma^2 \tag{30}$$

$$G_{Zeeman} = -BM_0\sigma \tag{31}$$

$$G_{elastic} = \frac{1}{2K}\left(\frac{V-V_0}{V_0}\right)^2 \tag{32}$$

$$G_{entropy} = -T(S_m + S_r) \tag{33}$$

$$G_{press} = P\left(\frac{V-V_0}{V_0}\right) \tag{34}$$

where $K$ is the compressibility coefficient and $N$ is the number of magnetic atoms per unit volume. By minimizing the equation 29 with respect to $\sigma$ and V, a modified (or generalized) expression of the Brillouin function can be obtained while the y function becomes [69-76]

$$y = \frac{1}{T}\left[3T_0\left(\frac{J}{J+1}\right)\sigma + \frac{g_J\mu_B J}{k}B + \frac{9}{5}\left(\frac{(2J+1)^4-1}{(2J+2)^4}\right)T_0\eta\sigma^3\right] \tag{35}$$

with the parameter $\eta = \frac{5}{2}\frac{[4J(J+1)]^2}{[(2J+1)^4-1]} NKk_B T_0 \beta^2$. This latter is of great importance since it defines the nature of the magnetic phase transition and involves the volume contribution. If $\eta > 1$, the transition is first order in nature while for η < 1, a second order phase transition occurs [53, 69-76]. Usually $\eta$ and $T_0$ parameters can be obtained by fitting theoretical thermomagnetic curves with experimental data [53, 63, 74]. Then the MCE in terms of ΔS and $\Delta T_{ad}$ can be calculated using the equations 21 to 25. For example, magnetization and MCE data of MnAs [53, 63, 74] that shows a typical first order magnetic transition are reported in Figures 4 and 5, respectively. For more complex systems as in the case of materials with itinerant electrons, other available models in the literature can be used. We particularly refer the interested reader to Refs. 64, 77-79.

### D. On the characterization of magnetocaloric materials

The characterization of magnetocaloric materials can be performed with the help of direct and indirect methods. For direct measurements of the magnetocaloric effect, the experiments are usually done in adiabatic conditions. For this purpose, the samples temperatures $T_I$ and $T_F$ corresponding to the change of the magnetic field from $B_I$ to $B_F$ must be determined accurately. Usually, at the beginning of each measurement, the initial temperature $T_I$ of the material is stabilized and then the magnetic field is changed from 0 to $B_F$. The corresponding adiabatic temperature change is then measured as the difference $\Delta T_{ad} = T_F - T_I$. The measurement accuracy depends on several factors such as, the thermal insulation of the sample, the thermal contact between the thermocouple and the sample, the equilibrium conditions and the magnetic field setting [52]. $\Delta T_{ad}$ can also be evaluated indirectly from specific heat measurements as a function of temperature in several constant magnetic fields. This technique enables to characterize the magnetocaloric effect in terms of both ΔS and $\Delta T_{ad}$ with the help of equation 11. It is worth noting that calorimetric measurements under magnetic fields are highly challenging. For this reason, only the specific heat for 0 T is frequently reported in the literature. In this case, $\Delta T_{ad}$ can be determined through the equation 25 by combining 0 T-specific heat data and obtained ΔS values via magnetic measurements (see following paragraphs). The needed full entropy for a given field can be expressed by S (B, T) = S (0, T) + ΔS (B, T), with S (0, T) is the full entropy at 0 T that can be calculated by $S(0,T) = \int_0^T \frac{C_P(0,T')}{T'} dT'$.

In order to measure $\Delta T_{ad}$, Levitin et al[80] have proposed an original technique based on adiabatic magnetization measurements. It consists to compare the magnetic field dependence of the magnetization under both adiabatic and isothermal processes. As a consequence of the temperature change under the effect of an external magnetic field, the adiabatic magnetization curve intersects the magnetic isotherms. The intersection point is utilized to identify the sample's final temperature and then $\Delta T_{ad}$ when magnetized in adiabatic conditions. However, due to the complexity of calorimetric measurements, the magnetocaloric effect is frequently reported in terms of $\Delta S$ that is deduced from isothermal magnetization measurements by using the numerical form of the well-known Maxwell relation (Eq. 9). This method enables a fast characterization of magnetocaloric materials. Since the magnetization measurements are realized at discrete magnetic fields and temperatures, the isothermal entropy change can be found through the numerical form of the Maxwell relation. In this case, the equation 9 becomes

$$\Delta S = \sum_i \frac{M_{i+1} - M_i}{T_{i+1} - T_i} \Delta B_i \qquad (36)$$

where $M_{i+1}$ and $M_i$ are the measured magnetizations in a field $B$, at temperatures $T_{i+1}$ and $T_i$, respectively. From a mathematical point of view, the isothermal entropy change is proportional to the area between two magnetic isotherms.

Based on this approach, and after a series of simple magnetization measurements, "huge" entropy changes have been reported in several materials which are presented as the "best refrigerants" for applications. However, the inadequate use of the Maxwell relation could results in spurious values of $\Delta S$, particularly when the considered materials present a first order magnetic phase transition [81-90]. In some cases, their phase transition is associated with a large hysteresis effect. Consequently, the material's magnetization strongly depends on the magnetic history which results in two different magnetic states for a certain value of the magnetic field. This means that the equilibrium state needed for the use of the Maxwell relation is not respected, which explain the overestimated values of $\Delta S$ reported in some materials such as $Mn_{1-x}Fe_xAs$ [85, 89, 90] and NiMnGa [91, 92]. By directly integrating the Maxwell relation between 0 and 5 T, the isothermal entropy change in $Mn_{1-x}Fe_xAs$ (for example) close to room-temperature was found to be as large as 325 J/kg K [89]. The latter is more than 30 times larger than that of gadolinium metal at about 294 K (-10 J/kg K under 5 T). This "colossal" value is mainly attributed to the inappropriate application of the Maxwell relation. In fact, the large hysteresis shown by $Mn_{1-x}Fe_xAs$ compounds lead to the coexistence of both ferromagnetic and paramagnetic phases in the temperature range close to $T_C$ [85, 89, 90]. In this case, only the paramagnetic phase contributes to the isothermal entropy change (MCE) when it is changed to a ferromagnetic phase (metamagnetic transition) under the effect of an external magnetic field. However, the direct application of the Maxwell relation also includes the ferromagnetic volume (Fig.6). Consequently, large parts of the area between two adjacent magnetic isotherms are unreasonably included in the integration process yielding to wrong estimation of the entropy change [85, 90, 93]. For example, we report in Figure 6 the magnetization isotherms for a typical first order magnetic transition material (MnAs) showing the coexistence of two magnetic phases (Ferro and Para). As plotted in the inset of Figure 6, the direct integration of the Maxwell relation largely overestimates ΔS values. A similar situation is frequently encountered in the Heusler's alloys. These materials usually present a first order magneto-structural transition (from AF to Ferro) which is accompanied by large hysteresis losses yielding to mixed antiferromagnetic and ferromagnetic states in the phase transition region [91-92]. In this case, only the antiferromagnetic phase accounts for the MCE when is transferred by the magnetic field to the ferromagnetic phase. In order to obtain realistic values for ΔS, several works suggested that the Maxwell relation must be integrated only within the field-induced metamagnetic phase transition region ($\Delta B_C$) [85-86, 88, 90].

$$\Delta S(T, \Delta B_C) = \int_{B_C - \frac{\Delta B_C}{2}}^{B_C + \frac{\Delta B_C}{2}} \left( \frac{\partial M}{\partial T} \right)_{H'} dB' \qquad (37)$$

$B_C$ is the critical magnetic field value within the transition zone. On the other hand, in metamagnetic materials more realistic values of ΔS can be obtained through the Clausius-Clapeyron (C-C) equation given by

$$\Delta S = -\Delta M \frac{dB_C}{dT} = -\Delta M \left( \frac{dT_T}{dB} \right)^{-1} \qquad (38)$$

where ΔM is the magnetization jump, $T_T$ is the transition temperature [85, 90]. This method directly links the magnetization jump and the corresponding entropy change. By using the C-C equation, the maximum value of ΔS in $Mn_{1-x}Fe_xAs$ was found to be only 26 J/kg K [85] instead 325 J/kg K initially reported in Ref.89. However, the C-C equation is more appropriate, particularly when the high magnetization phase tends to saturate after the metamagnetic phase transition. Otherwise, C-C values must be completed by integrating the Maxwell relation within the region outside the metamagnetic transition [88, 90].

It is worth noting that Caron et al [94] have proposed another approach for the evaluation of entropy change according to magnetization measurements even in materials displaying a large hysteresis effect. This method consists to eliminate the residual ferromagnetic volume by heating the considered material to the paramagnetic phase before each measurement. The proposed approach enables to obtain more reasonable values of ΔS. However, calorimetric measurements made at equilibrium conditions remain the best way for the evaluation of both ΔS and $\Delta T_{ad}$.

On the other hand, when measuring the MCE, another source of errors arises from the demagnetization effect caused by the magnetic materials' shape. In the literature, the magnetic and magnetocaloric properties are mostly reported with respect to the external magnetic field while neglecting the contribution of the demagnetization effect. When subjected to an external magnetic field, the measured magnetic substance creates in the opposite direction a demagnetizing field that cancel out a part of the applied external field. However, the internal magnetic field (or the local field) is the effective field acting on the magnetization and the specific heat, determining consequently the magnitude of MCE [95, 90]. Under an external magnetic field $H_0$, the local field in the sample is given by

$$H_{eff} = H_0 - N_d M \qquad (39)$$

where $N_d$ is the demagnetization factor that depends on the magnetic sample shape. The quantity $-N_d M$ represents the demagnetization field ($H_d$). For spherical forms, $N_d$ is equal to 1/3. Otherwise, the demagnetization factor can be determined by using Aharoni model for rectangular shapes [96] or other simplified approaches [62]. In Figure 7, we report the temperature dependence of the local magnetic field inside a sample of Gd under an external field of 1 T. As shown, the internal magnetic field markedly differs from the applied field particularly at low temperatures due to the large magnetization of the ferromagnetic phase. It is then extremely important to correct the reported MCE taking into account the demagnetization effect. This means that the magnetocaloric properties must be presented as a function of the effective magnetic field [95, 90].

Similarly to the demagnetization effect, the magnetocrystalline anisotropy can negatively impact the magnetocaloric effect in some magnetic materials [97]. This would consequently lower the thermodynamic performance of magnetic cooling devices. On account of the magnetic anisotropy usually shown by non-cubic magnetocaloric crystals, the MCE strongly depends on their orientation with respect to external magnetic field [97]. However, the MCE measurements are frequently performed by using polycrystalline samples. In this case, the obtained thermal effect rather corresponds to the average value of those resulting from the application of magnetic field along the easy, intermediate and hard-directions because of the arbitrary grain orientation [97]. For example, in a very recently reported work by Fries et al [97], it was found that the $Co_2B$ single crystal exhibits a maximum adiabatic temperature change of 0.9 K (at 425 K) under a magnetic field of 1.9 T applied along its easy-orientation, while it is only 0.65 K when the field is applied following the hard-direction. For the polycrystalline sample, a maximum adiabatic temperature change of 0.75 K is obtained in a similar magnetic field [97]. Thereby, in order to maximize the magnetocaloric effect in the AMR regenerators, the easy-axis of implemented magnetocaloric particles (grains) must be oriented along the direction of applied magnetic field.

## III. IMPLEMENTATION OF ADVANCED MATERIALS IN MAGNETIC COOLING

### A. Gd and related alloys

The rare earth elements and related alloys have attracted a worldwide interest due their utilization in several strategic domains such as microelectronic technologies, energy conversion and spintronic devices. The great interest given to rare earth alloys in magnetic refrigeration applications is mainly due to their excellent magnetocaloric properties near the ambient temperature, large magnetic moment, negligible hysteresis losses, high mechanical stability, possibility of use as refrigerants on a wide temperature range by tailoring their magnetic properties and their availability in the market. Additionally, the rare earth alloys enables to deal with several engineering requirements such as the possibility to obtain some specific shapes which is not permitted by the recently reported GMCE compounds. On the other hand, their localized magnetism allows the use of simplified theoretical models, namely the mean field theory to predict their performance in functional magnetic cooling machines. Their magnetic and magnetocaloric properties have been extensively studied from both practical and fundamental points of view [19, 52, 98-111].

The gadolinium metal (Gd) is the prototype material (reference) used in the most majority of room-temperature magnetic refrigerators [21]. Its magnetic and magnetocaloric properties are well known [52, 98]. At the Curie temperature $T_C$ = 294 K, Gd undergoes a second order magnetic phase transition from the low temperature ferromagnetic state to the paramagnetic phase. Taking into account the demagnetization effect, the maximum adiabatic temperature change $\Delta T_{ad}$ shown by Gd is about 3 K and 6 K under a magnetic field change from 0 to 1 T and 0 to 2 T, respectively. The corresponding entropy changes are about 3 J/kg K for 0-1 T and 5.5 J/kg K for 0-2 T. It is worth noting that the working magnetocaloric temperature range of Gd is limited close to room temperature where its MCE exhibits large values, on account of the magnetic phase transition taking place at 294 K. However, as reported in Figure 8, the cooling range of Gd can be markedly increased by chemical doping with other rare earths such as Tb and Dy for example [67, 68]. For this purpose, Smaili et al [67] have studied the magnetic and magnetocaloric properties of $Gd_{1-x}Dy_x$ alloys (with x = 0, 0.12, 0.28, 0.49 and 0.7) for Ericson-like magnetic refrigeration cycle tasks. They observed that the transition temperature can be drastically reduced from 293.5 K for Gd to 206.3 K for the $Gd_{0.3}Dy_{0.7}$ alloy. The isothermal entropy change was found to be practically unchanged with Dy doping up to x = 0.49. For x = 0.7 T, the obtained –ΔS (16 J/kg K for 7 T) exceeds that of Gd (12 J/kg K for 7 T) by about 33 %, particularly for sufficiently high magnetic fields. Based on these results, an optimum combination of $Gd_{1-x}Dy_x$ alloys in a multilayer has been proposed by the authors as a refrigerant operating over the temperature range 210-290 K. More details about multilayers (or composites) are reported in section VI.

Following, Hou et al [100] have investigated the adiabatic temperature change of $Gd_{1-x}Dy_x$ (x = 0 to 40 %) using commercial Gd and Dy with relatively low purity (up 99.8 %). When increasing the Dy content from 0 to 40 %, the Curie temperature was reduced from 288 to 245.5 K, while for a magnetic field of 1.2 T, the $\Delta T_{ad}$ at $T_C$ increases from 1.6 to 3.1 K, respectively. For Dy content between 27 and 40 %, the maximum $\Delta T_{ad}$ of $Gd_{1-x}Dy_x$ alloys obtained with low cost commercial elements is almost 3 K (for 1.2 T) which is comparable with that of high pure Gd (99.99 %). In the work by Balli et al [68], the $Gd_{1-x}Tb_x$ (x = 0, 0.3 and 0.5) alloys have been proposed as constituent materials for refrigeration over the temperature range 260-300 K. A good agreement was observed between the calculated Curie points of $Gd_{1-x}Tb_x$ using de Gennes model (see section II-C) and the corresponding experimental data. This means that with the help of de Gennes model [66], the transition temperature of each alloy can be determined and accordingly the desired temperature range as well as the needed contents. On the other hand, a multilayer material composed of $Gd/Gd_{0.7}Tb_{0.3}/Gd_{0.5}Tb_{0.5}$ (with the composition 55%/35%/10%) was proposed for application close to room-temperature. The optimum mass ratio of the constituent elements was calculated numerically and found to vary slightly with the magnetic field. The resulting entropy change (~ 4 J/kg K for 2 T) of the formed composite remains practically constant over the temperature range 260-300 K. The adiabatic temperature change of $Gd_{1-x}Tb_x$ alloys with x varying from 0 to 40 % was studied by Kastil al [105]. Under 1 T, the measured maximum values of $\Delta T_{ad}$ are about 2.5 K for all the studied samples, which is similar to that of pure Gd. The obtained transition temperature decreases from 294 K for Gd to 269 K for x = 0.4, confirming the early reported results by Balli et al [68].

More recently, the magnetic and magnetocaloric performance of $Gd_xHo_{1-x}$ (with x = 0.80, 0.91 and 1) alloys have been theoretically investigated in the framework of mean field theory and the de Gennes model [101]. The calculated entropy change of $Gd_xHo_{1-x}$ with x = 0.80, 0.91 and 1 is peaked at their respective transition points 265 K, 280 K and 293 K, respectively. This seems to be in good agreement with early reported experimental data [101]. The $-\Delta S$ was found to increase slightly with the decrease of Ho concentration. Under a magnetic field change from 0 to 2 T, $-\Delta S$ presents a maximum value of about 6 J/kg K. Based on numerical calculations, a multilayer refrigerant composed of $Gd_{0.80}Ho_{0.2}$, $Gd_{0.91}Ho_{0.09}$ and Gd was proposed with optimum mass ratios (under 2 T) equal to 0.24, 0.17 and 0.59, respectively. The composite is expected to work as refrigerant in the temperature range between 265 K and 293 K. Its performances in a regenerative Ericsson thermodynamic cycle were also analyzed by Xu et al [101]. The cooling energy shown by the composite (1008 J/kg under 2 T) exceeds largely that of individual $Gd_xHo_{1-x}$, while the calculated coefficient of performance (COP) reaches 9 for 2 T.

The magnetocaloric properties of the polycrystalline GdGa was investigated by Zhang et al [104]. This compound exhibit a low temperature ferromagnetic to paramagnetic transition around 183 K. The maximum values of $-\Delta S$ (4.81 J/kg K for 5 T) and $\Delta T_{ad}$ (4.43 K for 5 T) are about half than those of Gd. This can be mainly attributed to the non-magnetic character of gallium. Nevertheless, the broadening of $-\Delta S$ (T) profile enables a large relative cooling power (RCP). This latter was estimated to be 576 J/kg for a magnetic field variation of 5 T.

The potential use of Gd-based alloys as working refrigerants in an active magnetic regenerative cycle (AMR) [52] was also the subject of various studies [107-112]. Aprea et al [107] have performed a numerical analysis of an AMR refrigeration system with multilayer regenerators constituted of $Gd_{1-x}Tb_x$ alloys over the temperature range 275-295 K, and $Gd_{1-x}Dy_x$ alloys in the temperature range 260-280 K. The thermodynamic performances were found to markedly increase with the layer's number that can be obtained by varying the composition of $Gd_{1-x}R_x$ alloys. Comparing the COP value of an 8 layers AMR cycle with that of pertinent conventional compression-relaxation systems, the authors found that the AMR apparatus has an energetic performance larger than 63 %. On the other hand, $Gd_{1-x}R_x$ alloys were directly implemented in magnetic cooling systems leading to significant advances in terms of thermodynamic performances (see table 1). Rowe et al [109] have tested different multilayer regenerators composed of Gd, $Gd_{1-x}Tb_x$ and $Gd_{1-x}Er_x$, in an AMR apparatus using a magnetic field of 2 T and cycle frequencies of about 0.65 Hz. Different porous regenerators are made by using crushed particles of selected alloys with a mean diameter of about 0.35 mm. The best performance was obtained with the $Gd_{0.85}Er_{0.15}$- $Gd_{0.75}Tb_{0.26}$-Gd composite. The latter was able to deliver strong temperature spans up to 47 K which is about ten times the MCE peak of individual Gd or Gd-(Tb, Er) alloys, suggesting that efficient magnetic refrigerators could be built by simply using permanent magnets [109].

TABLE I. Implementation of $Gd_{1-x}R_x$ alloys in magnetic refrigerators.

| Research group | Device | B (T)[a] | Used Materials | Arrangement | $T_C$ (°C) | MCE (K) | Shape | Mass (kg) | f (Hz)[b] | Span (K)[c] | P (W)[d] | Ref. |
|---|---|---|---|---|---|---|---|---|---|---|---|---|
| Rowe et al | Linear | 2 (SC)[e] | Gd-Tb-Er | Composite (3 layers) | -8, 7, 22 | 5 (2T) | Particles | 0.135 | 0.65 | 49 | - | 109 |
| Zimm et al | Rotary | 1.5 (PM)[f] | Gd-Er | Composite (2 layers) | 10, 20 | - | Particles | - | 4 | 25 | 28 (14K) | 110 |
| Okamura et al | Rotary | 0.77 (PM) | Gd-Dy-Y | Composite (4 layers) | 2 to 10 | 1.5 (0.6T) | Spheres | 1 | - | - | 60 (1.1K) | 111 |
| Saito et al | Linear | 1.1 (PM) | Gd-Ho-Y | Composite (3 layers) | 0, 10, 15 | - | Spheres | - | 0.4 | 40 | - | 113 |

[a] is the strength of the magnetic field used by the magnetocaloric device during the magnetization-demagnetization process.
[b] is the operating frequency of the magnetic cooling machine.
[c] is the maximum obtained temperature difference between the hot and cold sources.
[d] is the cooling power produced by the magnetocaloric device.
[e] means that the used magnetic field source is a superconducting magnet.
[f] means that the used magnetic field source is based on permanent magnets.

In the study by Zimm et al [110], the performance of a rotary magnetic cooling device using $Gd_{1-x}R_x$ alloys have been reported and analysed. A layered bed consisting of spherical particles of Gd, diameter 425-500 μm and spherical particles of $Gd_{0.94}Er_{0.06}$, diameter 250-355 μm was used as refrigerant. The cooling process is achieved by the rotation of a wheel packed with the selected materials through a 1.5 T-permanent magnet [110]. With an AMR cycle frequency of 4 Hz, the Gd and Gd-Er layered bed showed large performances in comparison with the bed consisting entirely of Gd particles. For a temperature span of 14 K, the produced cooling power by Gd-GdEr refrigerant (28 W) is about twice larger than that obtained with Gd.

In the 0.77 T-rotary magnetic cooling device reported by Okamura et al [111], the AMR beds are constituted of four kinds of $Gd_{1-x}R_x$ alloy spheres presenting a diameter of 0.6 mm. The selected alloys are cascaded in the regenerator as follow: $Gd_{0.92}Y_{0.08}/Gd_{0.84}Dy_{0.16}/Gd_{0.87}Dy_{0.13}/Gd_{0.89}Dy_{0.11}$. A maximum cooling power of 60 W was obtained. The observed relatively low cooling power was attributed by the authors to some engineering issues such thermal losses and the low value of the magnetic field. However, with the improved versions of Okamura et al machine [112], a maximum cooling power of 540 W was reached for a temperature span of 0.2 K.

More recently, Saito et al [113] have tested several layered AMR-regenerators with $Gd_{1-x}R_x$ (R = Ho, Y) alloys aiming to reach cold temperatures in the sub-zero range. The experiments were carried out by using a 1.1 T-reciprocating magnetic cooling device where spherical particles of $Gd_{1-x}R_x$ are packed in a moving cylindrical regenerator that subjected to magnetization-demagnetization cycles. The used particles show a diameter of 500 μm. The heat transfer is performed by water or a 20 % glycol solution. The constituent alloys $Gd_{0.9}Ho_{0.1}$, $Gd_{0.95}Y_{0.5}$ and $Gd_{0.985}Y_{0.015}$ present a Curie temperature of about 0, 10 and 15 °C, respectively [113]. Their MCE in terms of the entropy change under 1 T is similar to that of Gd (about 3 J/kg K). When using as refrigerant the $Gd_{0.9}Ho_{0.1}/Gd_{0.95}Y_{0.5}/Gd_{0.985}Y_{0.015}$ multilayer in the proportions 10/3/10, respectively, the authors were able to generate a temperature span that exceeds 40 K, with a cycle frequency of 0.4 Hz. More interestingly, a cold temperature of -11 °C was attained, paving the way toward the commercialization of magnetic cooling. More details regarding the direct implementation of $Gd_{1-x}R_x$-based multilayers in magnetic cooling machines are given in table 1.

### B. $LaFe_{13-x}Si_x$-based compounds

The $La(Fe_xSi_{1-x})_{13}$ compounds present a ferromagnetic order in the concentration range 0.81 < x < 0.89 [27-32, 114-155]. Around $T_C$ = 200 K, they usually show a magnetic field-induced itinerant electron metamagnetic transition (IEMT) from paramagnetic to ferromagnetic state [27] resulting in a giant magnetocaloric effect (Fig.9). However, as shown in Figure 9, the direct implementation of these materials in room-temperature applications is not possible due to the low value of the Curie point. Therefore, the increase of $T_C$ toward room temperature without affecting their magnetocaloric properties is crucial before their utilization as refrigerants in functional devices. For this purpose, the hydrogen insertion in the $LaFe_{13-x}Si_x$ matrix enables to strongly shift $T_C$ toward room temperature while retaining a large magnetocaloric effect [26-27]. The insertion of other interstitial elements such as carbon and nitrogen also enhances the Curie point, but decreases drastically the magnetocaloric performance [31, 33, 34]. It was shown that the nitrogen absorption by $LaFe_{13-x}Si_x$ compounds drives drastically the magnetic phase transition from first to second order which strongly destroy the MCE [31]. On the other hand, when increasing the carbon content, the transition temperature can be shifted close to 260 K with reasonable magnetocaloric effect [33, 34]. Besides, it is difficult to use $LaFe_{13-x}Si_xC_y$ as refrigerants around 300 K, since a large amount of carbon is needed. This induces a significant decrease in the magnetocaloric performance and results in the appearance of secondary magnetic phases constituted of α-Fe [33, 34]. However, due to the strong Fe-Co exchange interaction, the substitution of a small content of Fe by Co in $LaFe_{13-x}Si_x$ drastically increases $T_C$ while retaining excellent magnetocaloric properties [28-30, 32].

Even though the hydrides $LaFe_{13-x}Si_xH_y$ (LaFeSiH) show a giant MCE, their mechanical brittleness as well as their chemical instabilities restrict their utilization in functional devices [114, 136]. In contrast, the more stable $La(Fe, Co)_{13-x}Si_x$ compounds (LaFeCoSi) have been more recently tested in magnetic cooling systems and promising results were obtained [126]. Since the cobalt is a strategic metal, it was shown in a previous work that by combining the cobalt and the interstitial carbon in $LaFe_{13-x}Si_x$ compounds, a large quantity of Co can be saved without affecting their magnetocaloric performances at room temperature [30].

Among $NaZn_{13}$ materials, the $LaFe_{13-x}Si_xH_y$ and $La(Fe, Co)_{13-x}Si_x$ compounds are currently the most utilized in magnetic refrigeration. Their magnetocaloric properties in terms of isothermal entropy and adiabatic temperature changes are summarized in Figure 10. As shown, the $LaFe_{13-x}Si_xH_y$ hydrides unveil large and almost unchanged entropy and adiabatic temperature changes of about 20 J/kg K and 6 K (under 2 T), respectively, over a wide temperature range. In contrast, the entropy change exhibited by $La(Fe, Co)_{13-x}Si_x$ becomes smaller for compounds with high $T_C$. Close to room temperature, their entropy change is usually about 8 J/kg K whereas the adiabatic temperature change is about 2 K/T (Fig. 10). It is worth noting that the $LaFe_{13-x}Si_x$ compounds have been widely explored in the past. In order to find more about their structural, magnetic and magnetocaloric properties, we refer the interested reader to several papers and reviews previously reported in the literature [114-157]. In this review, we mainly focus on their practical aspects.

Usually, $LaFe_{13-x}Si_x$ compounds crystallize in the cubic $NaZn_{13}$-type structure (1:13) with eight formula unit per crystal cell, where La occupies the 8a site and Fe goes on the 8b site. The 96i site is randomly shared by Si atoms and the rest of Fe [114-157]. Among the reported magnetocaloric refrigerants, $LaFe_{13-x}Si_x$-based materials are currently one of the most promising for applications at room temperature due to their good magnetocaloric properties and particularly the lower cost and the abundance of constituent elements compared to rare-earth- based alloys. However, even though the cost of needed starting elements is reasonable, the use of standard methods to prepare bulk $LaFe_{13-}$ $_x$Si$_x$ materials such as arc-melting and magnetic induction requires a long time annealing at around 1100 °C for several weeks to obtain products with high quality [114-157]. This could markedly increases the production cost of LaFe$_{13-x}$Si$_x$-based materials. Additionally, by using both techniques, it is challenging to keep the initial composition of starting elements during melting due to the evaporation of lanthanum. This usually affects the Curie temperature resulting in a large amount of α-Fe, which restrict the large scale production (in kilograms) of LaFe$_{13-x}$Si$_x$-based refrigerants. In this context, melt spinning method was found to dramatically reduce the annealing time and to result in a refined microstructure [156, 157]. By using this technique, Liu et al [156] were able to dramatically shorten the La(Fe, Co)$_{13-x}$Si$_x$ annealing time to only 1 hour. Unfortunately, the obtained ribbons cannot be directly implemented in functional devices due to their mechanical brittleness. Additionally, with melt spinning only a few quantities of NaZn$_{13}$ materials can be produced.

For large scale production of LaFe$_{13-x}$Si$_x$, the powder metallurgy was proven to be an effective appropriate preparation route [32, 129]. Using this method, Katter et al [32] have produced La(Fe, Co)$_{13-x}$Si$_x$ in kilogram quantities starting from commercial powders of Fe and Si which are mixed with LaH$_x$ and La-Fe-Co-Si powders. After sintering between 1333 K and 1433 K for 4 to 8 h under inert conditions, the resulting products show high densities of about 7.2 g/cm$^3$ and exhibit magnetocaloric performance comparable with values obtained with melting routes. On the other hand, the preparation process can be achieved by machining the obtained blocks in some specific shapes depending on the requirements of magnetic cooling devices. Applying this approach, parallel plates of La(Fe, Co)$_{13-x}$Si$_x$ were successfully prepared (Fig.11) by Vacuumshmelze company [32, 130] and provided to several research groups for test in their AMR-magnetic cooling prototypes. More recently, flakes of La(Fe, Co)$_{13-x}$Si$_x$-based materials were prepared in kilogram quantities by the strip casting method [118]. The obtained flakes showed a 95 vol. % of the NaZn$_{13}$-type phase, a negligible hysteresis and interesting magnetocaloric properties.

It is also worth noting that spherical particles of La(Fe, Co)$_{13-x}$Si$_x$ materials with the diameter ranging from 0.1 to 1.2 mm were successfully synthesized by using the rotating electrode process (REP) [131]. Their diameters can be controlled by the rotating electrode speed. The obtained spheres showed a large magnetocaloric effect close to room temperature. However, in order to obtain pure NaZn$_{13}$ phases, a heat treatment of the obtained spheres at 1323 K for more than 10 days is required when using the REP method, increasing markedly the cost of fabrication. In this context, Liu et al [129] have employed a different approach to prepare a rapidly solidified spherical particles of La(Fe, Co)$_{13-x}$Si$_x$. By using the drop-tube solidification technique and after a brief annealing at 1373 K for 1h, the authors [129] were able to obtain high purity regular spherical forms with the size ranging from 100 to 500 μm. The implementation of LaFe$_{13-x}$Si$_x$ particles as refrigerants in magnetic cooling systems enables a large specific surface area, enhancing the heat transfer in the regenerator. However, their use also results in a high pressure drop decreasing consequently the machine coefficient of performance (COP).

In recent years, several NaZn$_{13}$- based regenerators were experimentally tested. Their performances are summarized in Figure 12 and table 2. Zimm et al [152] have investigated the performance of La(Fe$_{1-x}$Si$_x$)$_{13}$H$_y$ hydrides in a 1.5 T -permanent-magnet rotary refrigerator. A bed consisting of irregular particles of La(Fe$_{0.88}$Si$_{0.12}$)$_{13}$H$_1$ with 250 to 500 μm size experiences a magnetization-demagnetization process thought the rotation of a wheel packed with other materials (Gd, Gd-Er) for comparison. At small temperature spans, the cooling capacity produced by La(Fe$_{0.88}$Si$_{0.12}$)$_{13}$H$_1$ was found to compares with that of Gd. In a following work, Russek et al [121] have explored and tested a bed packed with five layers of La(Fe, Si)$_{13}$H$_y$ with different Curie temperatures comprised between 12 °C and 22 °C. The used LaFeSiH materials present irregular forms with diameters changing from 0.25 mm to 0.4 mm and a porosity of 47 %. In a magnetic field change of 1.5 T, their isothermal entropy changes are comprised between 10 and 12 J/kg K. It was found that cooling powers higher than 400 W can be reached by using LaFeSiH particles. On the other hand, the layered LaFeSiH beds are able to produce a cooling power much larger than Gd at high temperature span. With a cycle frequency of 3.33 Hz and for a temperature span of 13.5 °C, 300 W of cooling power was generated by LaFeSiH particles, while only 150 W was produced by Gd. More recently, the implementation of LaFeSiH materials in a rotary magnetic refrigerator designed by Astronautics [47] produced a record cooling power higher than 2 kWs with a coefficient of performance superior to 2. The system that is described in Jacobs et al [47] uses a rotating permanent magnet employing a magnetic field of 1.44 T over twelve immobile regenerators consisting of several LaFeSiH-based spherical particles where the diameter is comprised between 177 and 246 μm. Each bed was packed with six layers of LaFeSiH presenting Curie temperatures ranging from 30.5 to 43 °C. The total mass of used LaFeSiH is 1.52 kg while the frequency of the AMR cycles is 4 Hz [47]. For a zero-temperature span, a maximum cooling power of 3042 W was reached, while it is 2090 W for a temperature span of 12 K which could be considered as the highest performance yet reported for a magnetic cooling machine. The coefficient of performance was found to be larger than 2 for temperature spans maintained below 10 K [47].

Although the $LaFe_{13-x}Si_xH_y$ hydrides have been successfully tested, their mechanical brittleness and the instability of hydrogen in the $LaFe_{13-x}Si_x$ matrix could restrict their utilization as refrigerants [114, 136]. These inconveniences explain the great attention paid to the implementation of $La(Fe, Co)_{13-x}Si_x$ materials in magnetic refrigeration systems (see Fig. 12 and table 2). However, the mechanical properties of $LaFe_{13-x}Si_xH_y$-based materials could be markedly improved by mixing them with the epoxy resign as demonstrated by Zhang et al [115]. For example, the $LaFe_{11.7}Si_{1.3}Co_{0.2}H_{1.8}$ bonded with 3 wt.% epoxy resign, shows a compressive strength of 162 MPa, exceeding that of bulk compound by 35 % while keeping a large magnetocaloric effect [115]. In the same way, Pulko et al [117] have investigated the mechanical and magnetocaloric characteristics of several epoxy-bonded LaFeCoSi plates. Their direct implementation in an AMR device employing a 1.15 T-magnetic field source generated a no-load temperature span of about 10 K. In addition, after several thousands of AMR cycles, the studied bonded plates showed no significant changes in their mechanical properties [117].

In order to compare different families of magnetocaloric materials, Engelbrecht et al [127] have studied the performance of various combination of $La(Fe, Co)_{13-x}Si_x$ in a simple AMR regenerator. For this purpose, flat plates of $LaFe_{11.06}Co_{0.86}Si_{1.08}$, $LaFe_{11.05}Co_{0.95}Si_{1.01}$ and $LaFe_{11.96}Co_{0.97}Si_{1.07}$ compounds with Curie temperatures of 3 °C, 13 °C and 16 °C, respectively, were directly implemented. In the used linear-AMR apparatus, the magnetic field is generated by a Halbach cylinder providing an average magnetic field of about 1 T. The considered plates with 0.9 mm thickness and 20 mm length can be arranged following different configuration to build regenerators with single and multilayer materials. By using a single material with $T_C$ around 16 °C, a no-load temperature span of 7.9 °C was reached for a utilization factor of 0.54, which is lower than that obtained by using gadolinium plates in similar conditions (about 9 °C). Noting that, the utilization factor (U) is defined as the rapport rate between the thermal capacity of the carrier fluid and that of the magnetocaloric refrigerant. The performance of a layered bed $La(Fe, Co)_{13-x}Si_x$ with Curie temperatures of 3 °C and 16 °C was also tested. The considered regenerator failed to produce a no-load temperature span larger than the single material. This was attributed by the authors to the fact that the two materials are not the appropriate combination [127]. In contrast, the configuration constituted of $La(Fe, Co)_{13-x}Si_x$ compounds with transition temperatures of 13 °C and 16 °C produce a no-load temperature span that slightly exceeds that of a single $La(Fe, Co)_{13-x}Si_x$ but still below that reached by gadolinium plates [127]. This contrasts with Balli et al data [126] showing that $La(Fe, Co)_{13-x}Si_x$ materials are enable to achieve a temperature span higher than Gd plates.

TABLE II. Implementations of LaFe$_{13-x}$Si$_x$-based materials in magnetic refrigerators.

| Research group | Device | B (T) | Used Materials | Arrangement | $T_C$ (°C) | MCE (K) | Shape | Mass (kg) | f (Hz) | Span (K) | P (W) | Ref |
|---|---|---|---|---|---|---|---|---|---|---|---|---|
| Jacob et al | Rotary | 1.44 (PM) | LaFeSiH | Composite (6 layers) | 30.5 to 43 | 4 (1.5T) | Particles | 1.52 | 4 | 18 | 3042(0K) 2090(12K) | 47 |
| Engelbrecht et al | Linear | 1 (PM) | LaFeCoSi | Composite (2 layers) | 13, 16 | 2 (1T) | Plates | 0.0713 | - | 8.5 | - | 127 |
| Legait et al | Linear | 0.8 (PM) | LaFeCoSi | Composite (4 layers) | 10 to 25 | 1 (0.8T) | Plates | - | - | 10.5 | - | 151 |
| Cheng et al | Linear | 1.5 (PM) | LaFeCoSiB | Composite (2 layers) | 6, 18 | 2.3 (1.5T) | Particles | 0.58 | 0.9 | 15.3 | - | 132 |
| Balli et al | Linear | 1.45 (PM) | LaFeCoSi | Composite (2 layers) | 7, 21 | 2 (1T) | Plates | - | - | 16 | - | 126 |
| Tusek et al | Linear | 1.15 (PM) | LaFeCoSi | Composite (4 layers) | 18.2 to 35 | 2 (1.2T) | Plates | 0.144 | - | 20 | - | 122 |
| Saito et al | Linear | 1.1 (PM) | LaFeCoSi | Single | 13 | - | Spheres | 0.1 | 0.3 | 22 | - | 120 |
| Saito et al | Linear | 1.1 (PM) | LaFeSiH | Single | 24 | 1 (0.8T) | Spheres | 0.1 | 0.3 | 20 | - | 120 |

In the work by Balli et al [126], a composite magnetocaloric material based on La(Fe, Co)$_{13-x}$Si$_x$ compounds was directly implemented in a linear preindustrial magnetic cooling machine and its performance was compared with that of Gd. The used magnetic refrigerator is composed of two parallel permanent magnets sources providing each one a magnetic field of about 1.45 T and two regenerators (see Fig.1). Each regenerator is divided into two separated parts. Consequently, when the first part of the regenerator is moved outside of the magnetic field region, the second part is automatically magnetized. This enables to drastically reduce the involved magnetic forces in the machine [126, 158]. In order to form a multilayer refrigerant, blocks of flat plates constituted of LaFe$_{\sim11.2}$Co$_{\sim0.8}$Si$_{\sim1.1}$ (50 %) with T$_C$ ~ 280 K and LaFe$_{\sim11.1}$Co$_{\sim0.9}$Si$_{\sim1.1}$ (50 %) with T$_C$ ~ 294 K were placed in the first regenerator of the cooling device.

Their maximum effective magnetocaloric effect is about 2.5 K/T and 2 K/T, respectively. The La(Fe, Co)$_{13-x}$Si$_x$ plates have a thickness of 1 mm and a width of 8 mm. the total width of the multilayer is 100 mm. The Gd plates (1mm*8mm*100mm) were placed in the parallel regenerator for comparison, in similar operating conditions. By using water as heat transfer fluid, the achieved maximum no-load temperature span is about 16 K for La(Fe, Co)$_{13-x}$Si$_x$ which slightly exceeds that obtained with Gd plates (14 K).

With the aim to optimize the performance of an AMR regenerator, Legait et al [151] have tested different La(Fe, Co)$_{13-x}$Si$_x$ based refrigerants in a reciprocating magnetic cooling machine based on permanent magnets. The used device is similar to that presented in Ref.127 and consists of a static AMR regenerator. The magnetization-demagnetization process is performed by a mobile Halbach-type magnet providing a magnetic field of 0.8 T. Four La(Fe, Co)$_{13-x}$Si$_x$ with different amount of Co resulting in Curie temperatures of 283, 288, 293 and 298 K were considered in the Legait et al work [151]. Under 1 T, their maximum entropy change is 8.1, 7.5, 7.2 and 6.8 J/kg K, respectively. The AMR regenerator contains a stack of parallel plates with 1 mm thickness, 22 mm width and 50 mm length. At first, only La(Fe, Co)$_{13-x}$Si$_x$ plates containing one material with transition point around 293 K were tested in different operating conditions leading to a maximum no-load temperature span of 8 K. However, with the regenerator containing four layered La(Fe, Co)$_{13-x}$Si, the temperature span was slightly improved to reach about 10.5 K, but remains lower than the Gd regenerator (11.5 K) [151]. The obtained result was attributed by the authors to the non-continuous T$_C$ of La(Fe, Co)$_{13-x}$Si$_x$ layers in the regenerator.

Tusek et al [122] have performed a comprehensive experimental study by using several AMR regenerators which consist of multi-layered La(Fe, Co)$_{13-x}$Si$_x$ refrigerants under various operating conditions aiming to compare the obtained results with the best Gd-based parallel plates AMR. The experiments were realized on a reciprocating magnetic cooling device using a Nd-Fe-B magnet assembly that provides a magnetic field of about 1.15 T. The cooling process is achieved by magnetizing and demagnetizing the involved magnetocalorics through a linear movement of the magnetic field source run by a pneumatic cylinder[122]. The heat transfer between the regenerators and the thermal sources is performed by a mixture of distilled water (66 %) and 33 % of a commercial automotive antifreeze based on ethylene-glycol. Furthermore, three AMR regenerators were layered with La(Fe, Co)$_{13-x}$Si$x$ presenting different Curie temperatures along the length of the AMR apparatus: two layered LaFeCoSi with T$_C$ = 18.2 and 23.8 °C, four layered LaFeCoSi with T$_C$ = 18.2, 23.8, 30 and 35 °C and seven layered LaFeCoSi with T$_C$ = 7.8, 10.8, 18.2, 23.8, 30, 35 and 39 °C. The dimensions of the AMR regenerator are 10mm*40mm*80mm. The LaFeCoSi flat plates are separated by a distance of 0.2 mm and show a thickness of 0.5 mm. Their maximum entropy and adiabatic temperature changes are about 5 J/kg K and 2 K under about 1.2 T. As reported in Ref.122, the resulting temperature span is very sensitive to the utilization factor and the AMR cycle frequency. At 0 Watts of applied cooling load, a maximum temperature span of about 20 K is obtained with the four and seven LaFeCoSi regenerators for a utilization factor of about 0.15. The regenerator with two layers of LaFeCoSi provides only a maximum temperature span of about 16 K in similar operating conditions that can be attributed to the narrow temperature range of its magnetocaloric effect. However, in both cases, the obtained span is lower than the Gd regenerator (23 K for U~ 0.3). This can be attributed to its sufficiently high MCE in terms of adiabatic temperature changes distributed on a large temperature span compared to LaFeCoSi materials as well as the better heat exchange in the Gd regenerator [122]. However, for small temperature spans, the LaFeCoSi-based regenerators could provide a larger cooling power if compared with the Gd-based AMR, which is mostly attributed by the author to the large values of the entropy change and the specific heat of LaFeCoSi compounds.

Cheng et al [132] have studied the refrigeration effect of $LaFe_{11.9-x}Co_xSi_{1.1}B_{0.25}$ (with x = 0.9 and 0.82) compounds (LaFeCoSiB) in a reciprocating magnetic cooling device and the obtained data were compared with the Gd metal. Different tests were carried out with the help of a linear magnetic refrigerator based on a Halbach type Nd-Fe-B permanent magnet that provides a magnetic field of 1.5 T. In order to prevent oxidation of $LaFe_{11.9-x}Co_xSi_{1.1}B_{0.25}$, a mixture solution of $Na_2MoO_4$, $Na_3PO_3$, $NaCr_2O_7$ and $Na_2SiO_3$ was selected as heat transfer fluid. The MCE is induced by linearly moving the magnet. The AMR cycle frequency is 0.9 Hz and the heat transfer fluid follow rate is 5 ml/min. The tested $LaFe_{11.9-x}Co_xSi_{1.1}B_{0.25}$ were prepared by the magnetic induction method and exhibit Curie temperatures of 291 and 279 K for x = 0.9 and x = 0.82, respectively. Both compounds present a maximum adiabatic temperature change of about 2.3 K under a magnetic field of 1.5 T. Two different regenerators made of a single $LaFe_{11}Co_{0.9}Si_{1.1}B_{0.25}$ (580 g) and a composite of $LaFe_{11}Co_{0.9}Si_{1.1}B_{0.25}$ (390 g) and $LaFe_{11.08}Co_{0.82}Si_{1.1}B_{0.25}$ (190 g), with irregular particles (size from 0.42 to 0.86 mm) were tested. The regenerator constituted of $LaFe_{11}Co_{0.9}Si_{1.1}B_{0.25}$ particles enables to reach a maximum temperature span of about 12.7 °C being slightly lower than that of 785 g of Gd particles (14.9 °C). However, in the condition of same mass (580 g), $LaFe_{11}Co_{0.9}Si_{1.1}B_{0.25}$ particles provide a maximum temperature span (12.7 °C) that is 1.57 larger than that of Gd particles (8.1 °C). On the other hand, the implementation of LaFeCoB-based composite enables to improve the device performance leading to a maximum temperature span of 15.3 °C [132].

More recently, Saito et al [120] have explored the cooling properties of spherical particles composed of LaFe$_{13-x}$Si$_x$- based materials with diameters changing from 0.2 to 1.2 mm by using an AMR device. The obtained results were discussed in the framework of those generated by Gd-based alloys particles. The considered LaFe$_{13-x}$Si$_x$ compounds were synthesized by using the rotating electrode process and have Curie temperatures of 24 °C and 13 °C for La(Fe$_{0.86}$Si$_{0.14}$)$_{13}$H$_{1.2}$ and La(Fe$_{0.85}$Co$_{0.07}$Si$_{0.08}$)$_{13}$, respectively. Their maximum entropy change is about 3 J/kg K in an external magnetic field change of 0.8 T. In a similar magnetic field, the peak value of the adiabatic temperature change is about 1 °C in the case of La(Fe$_{0.86}$Si$_{0.14}$)$_{13}$H$_{1.2}$. The materials were packed into a cylindrical regenerator which is magnetized and demagnetized by linearly moving it inside and outside of an approximately 1.1 T magnetic field source. The water is used as heat transfer fluid while the AMR-cycle frequency is 0.3 Hz. In order to avoid oxidation and corrosion phenomena, the LaFe$_{13-x}$Si$_x$-based particles were coated with copper. For both La(Fe$_{0.86}$Si$_{0.14}$)$_{13}$H$_{1.2}$ and La(Fe$_{0.85}$Co$_{0.07}$Si$_{0.08}$)$_{13}$ regenerators, a maximum no-load temperature span of about 22 °C was reached being 10 °C lower than that obtained with Gd particles. This can be explained by the low value of $\Delta T_{ad}$ caused by the greater specific heat of LaFe$_{13-x}$Si$_x$ materials. However, in order to understand the effect of specific heat on the cooling properties, Saito et al [120] have also performed measurements with heat-load. They found that LaFe$_{13-x}$Si$_x$ materials show better heat-load properties when compared with Gd-based regenerators.

In a recently reported work, Bez et al [159] have studied the performance of epoxy-bonded La(Fe, Mn, Si)$_{13}$H$_z$ regenerators in their linear 1.1 T-AMR device described in Ref. 127. Both single and double-layered regenerators were tested. The bonded regenerators are constituted of irregular particles with sizes ranging from 250 to 500 μm and show a porosity of 55 %. The water mixed with a small amount of anticorrosion additives was utilized for the heat transfer between the hot and cold sources. The utilization of a 95 g double layer regenerator (T$_C$ = 23 and 26.6 °C) with 2 wt. % epoxy enables to generate a no-load temperature span that exceed 13 K for a low AMR frequency of 0.13 Hz. Based on their experimental tests, the authors suggested that 2 wt.% of epoxy maximizes the temperature span while retaining a high mechanical stability [159].

### C. MnFeP$_{1-x}$As$_x$ -based compounds

The phosphide-arsenide MnFeP$_{1-x}$As$_x$-based compounds [43, 160-176] belong to a wide family of pnictides with MM'X formula (M, M' = 3d or 4d metals and X = P, As, Ge, Si) that usually crystallize in the hexagonal Fe$_2$P type crystalline structure. The Fe$_2$P crystallizes in the hexagonal phase with space group P-62m. Its crystallographic structure exhibits two different metal sites, a pyramidal Fe (3g) with five P as nearest neighbours (NN) and a tetrahedral Fe (3f) with four P as NN. In MnFeP$_{1-x}$As$_x$ series, Mn atoms preferentially occupy the 3g site, while Fe atoms go to 3f site. This family of compounds whose fundamental properties were former studied in details, has attracted a great interest during last fifteen years due to its large magnetocaloric properties and low cost [43, 170, 175]. In 2002, a giant magnetocaloric effect and tunable magnetic properties were pointed out by Tegusi et al [43] in MnFeP$_{1-x}$As$_x$ materials leading to cover a large working temperature range only by varying the As/P ratio.

Although MnFeP$_{1-x}$As$_x$ materials unveil a large MCE around room temperature, the presence of toxic elements such as As drastically restricts their utilization as refrigerants in commercial devices. On the other hand, the difficulty to prepare MnFeP$_{1-x}$As$_x$ in large quantities due to the high vapour pressure of As as well as their large hysteresis constitute an additional obstacle to their implementation [19]. For this purpose, several efforts were made in order to eliminate the As element [163-166, 168, 170-172] leading to the more interesting systems MnFe(P, Si, Ge) where the magnetocaloric properties are summarized in Figures 13 and 14. In Trung et al [172], the magnetic and magnetocaloric properties were tailored by tuning the compositions of P/Ge and Mn/Fe in the Mn$_{1.1}$Fe$_{0.9}$P$_{1-x}$Ge$_x$ and Mn$_{2-y}$Fe$_y$P$_{0.75}$Ge$_{0.25}$ compounds, respectively. It was found that when increasing the Ge content, the Curie temperature increases from about 260 K for x = 0.19 to about 290 K for x = 0.22 (Fig.13-a) while the thermal hysteresis decreases from 6 to 4 K, respectively. On the other hand, the increase of the Mn amount in Mn$_{2-y}$Fe$_y$P$_{0.75}$Ge$_{0.25}$ enables to reduce both the transition temperature and the thermal hysteresis. For y changing from 0.84 to 0.8, T$_c$ varies from about 320 to 300 K (Fig. 13-c) while the thermal hysteresis is suppressed for y = 0.8. At room temperature, both compounds Mn$_{1.1}$Fe$_{0.9}$P$_{1-x}$Ge$_x$ ( x = 0.22) and Mn$_{2-y}$Fe$_y$P$_{0.75}$Ge$_{0.25}$ (y = 0.8) present a large -ΔS$_{max}$ of about 20 and 12 J/kg K under a magnetic field change of 2 T, respectively.

Later, Dung et al [166] have shown that by varying the Mn/Fe ratio in Mn$_x$Fe$_{1.95-x}$P$_{0.50}$Si$_{0.50}$, a small hysteresis lower than 1 K can be obtained while keeping excellent magnetocaloric properties, opening the way for the implementation of these materials in functional devices. Wada et al [170] have demonstrated that the increase of the Ru content in Mn$_{1.2}$Fe$_{0.8-z}$Ru$_z$P$_{0.5}$Si$_{0.5}$ compounds decreases both the Curie temperature (Fig. 13-d) and the thermal hysteresis. When increasing the Ru content from 0 to 0.15, T$_C$ is decreased from about 320 to 276 K and the thermal hysteresis is significantly reduced from 4.2 to 1.8 K, respectively. The maximum isothermal entropy change remains approximately constant for z between 0 and 0.15 being about 13 J/kg K in a magnetic field change of 2 T. The indirect estimation of the adiabatic temperature change of $Mn_{1.2}Fe_{0.7}Ru_{0.1}P_{0.5}Si_{0.5}$ gives rise to a maximum value of 4.3 K in a field change of 2 T [177]. With $Mn_{1.2}Fe_{0.75-y}Ni_yP_{0.5}Si_{0.5}$ compounds, the increase of Ni amount markedly lowers (Fig. 13-b) the transition temperature [170]. As reported by Wada et al [170], the thermal hysteresis in $Mn_{1.2}Fe_{0.75-y}Ni_yP_{0.5}Si_{0.5}$ could be suppressed for y = 0.1. In the concentration range $0 \leq y \leq 0.1$, the maximum entropy change is comprised between 8 and 12 J/kg k under the field change of 2 T (Fig. 13-b). For both Ni and Ru compounds, the values of the refrigerant capacity are in the range 180-200 J/kg under 2 T [170].

Recently, Yibole et al [165] have measured the magnetocaloric effect of $Mn_xFe_{1.95-x}P_{1-y}Si_y$ in terms of the adiabatic temperature change, $\Delta T_{ad}$. In order to optimize the MCE of MnFe(P, X), the $\Delta T_{ad}$ was firstly reported for different composition of $Mn_xFe_{1.95-x}P_{1-y}Si_y$ (y = 0.5). Once again, the transition temperature decreases with increasing the Mn amount. For x changing from 1.24 to 1.28, $T_C$ decreases from around 278 to 268 K (Fig.14-a). Among these compounds, the material with x = 1.24 shows the largest isothermal entropy change (13.5 J/kg K for 2 T). However no significant difference is observed concerning the maximum value of $\Delta T_{ad}$, being about 2 K in the field of 1.1 T for all the compositions (Fig.14-c). Based on magnetic and magnetocaloric considerations, the authors opted then for x = 1.25 as the optimum composition [165]. Following, the $\Delta T_{ad}$ of $Mn_{1.25}Fe_{0.7}P_{1-y}Si_y$ was explored. The obtained data demonstrate that the decrease in the Si amount from y = 0.52 to 0.49 enhances the thermal hysteresis while reducing $T_C$ from about 302 to about 278 K, respectively (Fig. 14-b). The maximum entropy change was found to be about 10 and 12 J/kg K for y = 0.52 and 0.51, respectively in the field change of 2 T. For y = 0.5 and 0.49, $-\Delta S_{max}$ is about 15 J/kg K in a similar magnetic field (Fig. 14-b). Under a magnetic field change of 1.1 T, the $Mn_{1.25}Fe_{0.7}P_{1-y}Si_y$ compositions exhibit a maximum $\Delta T_{ad}$ of about 2 K (Fig. 14-d), which is similar to that reported in $Mn_xFe_{1.95-x}P_{1-y}Si_y$ (y = 0.5) compounds (Fig.14-c) [165]. On the other hand, the $Mn_{1.2}Fe_{0.8}P_{0.75}Ge_{0.25}$ which presents a phase transition close to $T_C$ = 282 K and investigated in Ref.165 unveils a maximum adiabatic temperature change of about 1.8 K under 1.1 T, in contrast with its large value in terms of the maximum entropy change (10.1 J/kg K for 1 T). For more information about recent developments concerning the new generation of Mn-based intermetallics, we refer the interested reader to Refs. 165-167, 170, and 172-174.

As reported in the literature, the MnFe(P, Ge, As, Si) materials are usually prepared using several techniques such as, melt-spinning method, ball-milling technique and spark plasma sintering (SPS) technology [167]. In order to upscale these materials to industrial levels, BASF Company has proposed a method for generating a giant magnetocaloric effect in MnFePSi compounds [178]. On the other hand, Wada et al [170] have successfully scaled up the production of Mn-based compounds in large quantities with different shapes. The constituted elements were first mixed by using the ball milling technique and then sintered in a furnace under argon atmosphere. Based on the composition $Mn_{1.2}Fe_{0.735}Ru_{0.065}P_{0.45}Si_{0.55}$, the authors were able to produce plate-type materials up to 250 g and rod-type materials up to 700 g [170]. The obtained results demonstrate excellent reproducibility of the magnetic and magnetocaloric properties [170].

In a recent work, BASF has successfully produced the compounds $Mn_xFe_{2-x}P_{1-y}Si_y$ by gas atomization process on a 2 kg level [168]. The resulting spherical particles were subjected to a heat treatment in an Argon atmosphere at temperatures from 800 to 1200 °C for several hours [168]. For y = 0.53, the transition temperature of $Mn_xFe_{2-x}P_{1-y}Si_y$ was tailored and shifted from about 305 K to about 255 K by increasing the Mn content. The synthesized spheres present an average size of 100 μm and reveals maximum entropy changes between 12 and 18 J/kg K under 1.5 T. Their adiabatic temperature change was found to be about 1.8 to 2 K in the magnetic field change of 1.1 T. In a following work, more stable and porous layered –regenerators constituted of $Mn_xFe_{2-x}P_{1-y}Si_y$-based spheres were built by bonding them together using epoxy and subsequent heat treatment in temperatures ranging from 100 to 200 °C [179].

It is worth noting that until now only few studies were devoted to the direct test of $MnFeP_{1-x}As_x$ in magnetic cooling devices, which contrast to $LaFe_{13-x}Si_x$ compounds for example (see section III-B). However, their implementation is expected to markedly increase in the forthcoming years due to the recent development in terms of preparation techniques and magnetocaloric performances. In Campbell et al [169], regenerators composed of three, six and eight layers of $MnFeP_{1-x}As_x$ particles with different Curie temperatures have been tested in the 1.1 T-rotary magnetic refrigerator developed in the University of Victoria [180]. The $MnFeP_{1-x}As_x$ particles present irregular forms with diameter going from 300 to 425 μm. It was found that the temperature span increases significantly when increasing the number of $MnFeP_{1-x}As_x$ layers, confirming the already reported calculation on the AMR cycle [169]. By using 150 g of eight layered $MnFeP_{1-x}As_x$ with $T_C$ = 2.1, 6.1, 10, 15, 18, 22.4, 26.3 and 30 °C, a maximum no-load temperature span of 32.2 °C was reached for an AMR cycle frequency of 0.7 Hz. With three layered $MnFeP_{1-x}As_x$ (58 g) with $T_C$ = 14, 18 and 22 °C, a no-load temperature span of only 14.4 °C was achieved for an AMR cycle frequency of 0.8 Hz. As demonstrated by Campbell et al [169], the low cost $MnFeP_{1-x}As_x$ materials show a great potential for application in magnetic refrigeration. What remains now is to directly evaluate the performance of As-free Mn- based materials such as MnFe(P, Si, Ge), in magnetic cooling devices. In a recently reported work [168], the implementation of a $Mn_xFe_{2-x}P_{1-y}Si_y$ material with $T_C = 294$ K in an AMR magnetic cooling system resulted in a no-load temperature span of 10 K. This initial result is very encouraging and constitutes an important step toward the utilization of $Fe_2P$-type materials as refrigerants.

In order to understand the irreversibilities associated with the first order magnetic transition usually shown by $MnFeP_{1-x}Si_x$, a single-layer of $MnFeP_{1-x}Si_x$ particles (50.6 g) has been more recently studied by Govindappa et al following the heating and cooling procedures [181]. The considered particles unveil irregular forms with diameter changing from 300 to 425 μm, and a maximum magnetocaloric effect of about 1.7 K under 1.1 T. The performance measurements were carried out at no-load conditions using the magnetic refrigerator described in Ref. 180. The AMR cycle operating frequency is 1 Hz while the heat exchange is performed by using a mixture of water and ethylene in a volume fraction of 80/20 % [181]. The results show a meaningful difference between the heating and cooling processes maximum temperature span as a function of the rejection temperature [181]. For example, around 34 °C a temperature span of 10.4 °C is obtained with the heating process and only 7.3 °C is reached with the cooling process. This underlines the negative impact of hysteretic effects on the AMR-cycle performance.

### D. Implementation of oxides in magnetic cooling systems

In addition to excellent magnetocaloric properties, the considered magnetocaloric materials must deal with additional series of requirements before their implementation as refrigerants in functional devices, such as, high electrical resistance, mechanical stability, safe constituent elements and high chemical stability. In contrast with the intermetallic compounds, the manganese oxides could largely answer these practical restrictions, which compensate for their relatively moderate magnetothermal effects [19, 46, 182-184]. Particularly, the manganites with general formula $R_{1-x}A_xMnO_3$ (R = lanthanide, A = divalent alkaline earth) have attracted a wide interest which is due to their interesting levels of the MCE close to room-temperature as well as to the possibility of tailoring their magnetic and magnetocaloric properties by doping the rare earth and manganese sites [46, 182-184]. In fact, the physical properties of such materials are usually controlled by the super-exchange coupling involving the Mn-O-Mn bond which could drastically affected by any structural or/and electronic changes caused by doping. The manganites-based materials have been widely investigated in the literature and their structural, magnetic and magnetocaloric properties are well known. For more details, we refer the interested reader to the recently reported works in Refs. 19, 46, and 182-184. In this work, we mainly focus on their direct implementation in functional magnetic cooling systems (see Fig.15 and table 3). From a practical point of view, the $La_{2/3}(Ca, Sr)_{1/3}MnO_3$ based materials are considered as one of the best candidates among the oxide magnetocalorics due to their large magnetization and high transition temperature. Although this kind of materials unveils a relatively low adiabatic temperature change when compared with reference magnetocalorics such as LaFeCoSi and Gd, their large specific heat enables an entropy change similar to that of Gd metal [182, 183, 184].

Bahl et al [185] have recently explored the performance of a multilayer refrigerant composed of $La_{0.67}Ca_{0.2925}Sr_{0.0375}Mn_{1.05}O_3$ (LCSM-1) and $La_{0.67}Ca_{0.2850}Sr_{0.0450}Mn_{1.05}O_3$ (LCSM-2) compounds in an AMR setup. Both materials were synthesized by using the spray pyrolysis technique. The resulting powders were subjected to a heat treatment performed at 1273 K for 2 h and then formed in a slurry with the help of a mixture of methylethylketone and ethanol, polyvinyl pyrolidone and polyvinyl butyral[185]. More details concerning the manufacturing of the used LCSM plates can be found in Ref. 185. In order to form the composite refrigerant, 28 platelets with a total mass of 51.1 g were stacked along the direction of the heat carrier fluid (water with 20 % of commercial ethylene glycol). Each platelet is constituted of similar content of LCSM-1 and LCSM-2 presenting the size 40mm*25mm*0.3mm. The Curie temperature of LCSM-1 and LCSM-2 platelets was found to be 277 and 282 K, respectively. The entropy and adiabatic temperature changes of the two platelets were measured in an applied magnetic field of 1 T. Close to $T_C$, $-\Delta S$ present maximum values of 3.7 J/kg K for LCSM-1 and 3.5 J/kg K for LCSM-2. The corresponding $\Delta T_{ad}$ are 1.3 K and 1.17 K, respectively. The LCSM-1/LCSM-2 multilayer refrigerant has been directly tested in a reciprocating AMR device using a Halbach-type permanent magnet structure producing a magnetic field of 1.1 T. The used device is well described in Ref. 127. For a utilization factor of 0.4 and a fluid rate of 1.32 g/s, a temperature span of 9.3 K was obtained at a hot source temperature of 283.8 K, being 7.5 times larger than the MCE presented by LCSM-1 and LCSM-2 compounds. On the other hand, the reached span is similar to that generated by Gd, demonstrating the high potential of manganites as refrigerants in magnetocaloric devices.

TABLE III. Implementation of oxides in magnetic refrigerators.

| Research group | Device | B (T) | Used Materials | Arrangement | $T_C$ (°C) | MCE (K) | Shape | Mass (kg) | f (Hz) | Span (K) | Ref. |
|---|---|---|---|---|---|---|---|---|---|---|---|
| Bahl et al | Linear | 1.1 (PM) | LaCaSrMnO | Composite (2 layers) | 4, 9 | -1.2 (1T) | Plates | 0.0511 | - | 9.3 | 185 |
| Engelbrecht et al | Linear | 1 (PM) | LaCaSrMnO | Single | 23 | 1 (1T) | Plates | 0.0341 | - | 5.1 | 127 |
| Guillou et al | Linear | 0.8 (PM) | PrSrMnO | Single | 22 | 1.1 (1T) | Plates | - | 0.18 | 5 | 186 |

In addition to Gd and LaFeCoSi-based materials where the performance are discussed in section III-B, Engelbrecht et al [127], have also tested the $La_{0.67}Ca_{0.26}Sr_{0.07}Mn_{1.05}O_3$ (LCSM) oxide in the same device described in Ref. 127. The used material was synthesized by the tape casting method. The obtained plates have a length of 40 mm following the direction of the heat transfer fluid circulation, a width of 25 mm and a thickness of 0.3 mm. The total mass of the used LCSM is 34.1 g. On the other hand, LCSM presents a Curie temperature of 23 °C. Under a magnetic field change of 1 T, the maximum values of its entropy and adiabatic temperature changes are about 17 kj/m$^3$ and 1 °C, respectively [127]. It was found that the generated temperature span is slightly dependent on the AMR cycle time, whereas it is highly sensitive to the utilization factor [127]. In the ambient temperature of 25 °C, a maximum no-load temperature span of 5.1 °C is achieved for an optimum utilization factor of approximately 0.55, being lower than the reached span when using Gd (-10 °C) and a single LaFeCoSi material (-8 °C). This can be mainly explained by the fact that the LSCM has a lower magnetocaloric effect (1 °C/T) if compared with Gd metal (3.2 °C/T) and LaFeCoSi (1.8 °C/T) [127].

It is known that the manganese perovskites $La_{2/3}(Ca, Sr)_{1/3}MnO_3$ are one of the best magnetocaloric oxides working at the room-temperature range. However, the magnetocaloric $Pr_{1-x}Sr_xMnO_3$ compounds are also very promising from a practical point of view. In comparison with $La_{2/3}(Ca, Sr)_{1/3}MnO_3$, the $Pr_{1-x}Sr_xMnO_3$ compounds exhibit similar magnetocaloric properties. In addition, the limited number of constituent elements in the $Pr_{1-x}Sr_xMnO_3$ enables a better control of the magnetic properties and the synthesis process, particularly during the large scale production step (kilograms) of selected compositions [186]. In the work by Guillou et al [186], the performance of a regenerator containing the $Pr_{0.65}Sr_{0.35}MnO_3$ compound was carried out using a 0.8 T-AMR test bench. Firstly, 0.6 kg (powder) of the selected compound was obtained through the solid state reaction. The appropriate dimensions of the desired plates (25*20*1 mm³) were obtained by cutting the compacted powder (blocks) using a circular saw [186]. In order to cover the regenerator length (50 mm), the plates were stacked along the heat transfer fluid flow direction, two by two [186]. Before its direct implementation, the physical and magnetocaloric properties of $Pr_{0.65}Sr_{0.35}MnO_3$ were characterized in terms of the entropy and adiabatic temperature changes, the thermal conductivity and the electrical resistivity. The material shows a Curie temperature (295 K) similar to that of the reference benchmark metal that is Gd ($T_C$ = 294 K). For a field variation of 1 T, a maximum entropy change of 2.3 J/kg K was reported in $Pr_{0.65}Sr_{0.35}MnO_3$. The corresponding adiabatic temperature change was found to be about 1.1 K, which is much lower than that exhibited by Gd (3 K /T) [90]. This is mainly due to the large specific heat of $Pr_{0.65}Sr_{0.35}MnO_3$ as reported in Ref. 186. On the other hand, it was also reported that the thermal conductivity of $Pr_{0.65}Sr_{0.35}MnO_3$ is about 6 times lower if compared with that of Gd [186] which could limit the heat transfer during the AMR cycle. This is usually a common point of a wide number of manganese perovskites. However, the large electrical resistance shown by $Pr_{0.65}Sr_{0.35}MnO_3$ could compensate for its lower thermal conductivity by minimizing the thermal losses caused by the eddy currents during the magnetization-demagnetization process.

Starting from an ambient temperature around 20 °C, the $Pr_{0.65}Sr_{0.35}MnO_3$ regenerator was able to provide a no-load temperature span of about 5 K for a frequency of 0.18 Hz , a flow rate of 0.5 mL/s and a utilization factor of 0.14, which is 5.6 times larger than the MCE at 0.8 T. On the other hand, in similar conditions the generated span is slightly lower than that of Gd (6.3 K) but compares well with that provided by the $La_{0.67}Ca_{0.26}Sr_{0.07}Mn_{1.05}O_3$ material (5.1 K) where the performance are reported in Ref. 127. However, when increasing the flow rate up to 1 mL/s, the obtained temperature span is only 4 K for $Pr_{0.65}Sr_{0.35}MnO_3$ which is less than 50 % of that shown by Gd (9.8 K). This reflects the significant difference between the physical properties of Gd and $Pr_{0.65}Sr_{0.35}MnO_3$ materials, such as the thermal conductivity, the specific heat and the adiabatic temperature change [186]. In a following paper, Legait et al [151] have studied the performance of the $Pr_{0.65}Sr_{0.35}MnO_3$ compound using the same AMR-device as in Guillou et al work [186] but with a wide range of working conditions, aiming to define the optimum operating parameters. Unfortunately, this investigation has failed to improve the performance of $Pr_{0.65}Sr_{0.35}MnO_3$ in an AMR cycle since the obtained maximum span (about 5 K) is similar to that reported in Ref. 186.

## IV. MAGNETOCALORIC MATERIALS AND STABILITY ISSUES

One of the most advantages favouring the magnetocaloric oxides against the intermetallics is their high resistance to corrosion and oxidation phenomena, which was confirmed in Ref.186. In fact, the heat transfer between the regenerator part and the end sources in the magnetic cooling systems is performed with the help of a moving carrier fluid. Thanks to its excellent thermal properties such as large specific heat, the water based fluids are usually used for the heat transfer. Concerning the oxides, Guillou et al [186] have studied the resistance of $Pr_{0.65}Sr_{0.35}MnO_3$ to corrosion. After immersing this manganite in water for different periods, its magnetocaloric properties in terms of entropy changes remain practically unchanged even after 1 years and half [186]. However, in contact with water the magnetocalorics and particularly the intermetallic based materials oxide easily resulting in the degradation of the thermodynamic performance and the working life of magnetic refrigeration devices, as shown in Figure 16-a. In addition, the magnetocaloric materials will be frequently in contact with air during the production, storage and recycling phases, which also favors their oxidation. In order to address these issues, several works regarding the chemical stability of magnetocaloric materials were recently reported in the literature [186-195].

In the pioneer work by Z. Y. Zhang et al [187], the chemical stability of commercial gadolinium in the presence of water was investigated. For this purpose, the gadolinium was immersed in a NaOH solution for long time. The obtained results demonstrate that no corrosion or weight losses were observed making from NaOH solution a good potential candidate as heat exchange media. What remains now is the study of its thermal properties. In the same way, M. Zhang et al [188] have explored the corrosion behaviour and its effect on the magnetic and magnetocaloric properties of $La(Fe, Co)_{13-x}Si_x$ compounds by using different techniques such as X-ray diffraction, scanning electron microscopy, X-ray photoelectron spectroscopy, magnetization measurements and weight loss method. The corrosion investigations were performed in distilled water using the compound $La(Fe_{0.94}Co_{0.06})_{11.7}Si_{1.3}$. It was found that the corrosion of $La(Fe_{0.94}Co_{0.06})_{11.7}Si_{1.3}$ is due to the electrochemical inhomogeneity of its surface. The final substances of corrosion on sample surface were identified as $La_2O_3$, $\gamma$-Fe(OOH), $Co(OH)_2$ and $H_2SiO_3$. It is worth noting that after 15 days corrosion in distilled water, the Curie temperature of the $La(Fe_{0.94}Co_{0.06})_{11.7}Si_{1.3}$ compound remains practically constant around 290 K. However, its maximum entropy change under the field variation of 2 T was reduced by about 16 % only after two weeks immersion in distilled water. This is explained by the fact that, the entropy change depends of the magnetocaloric phase. Consequently, the decrease of the La(Fe$_{0.94}$Co$_{0.06}$)$_{11.7}$Si$_{1.3}$ phase mass due the corrosion effect results in the reduction of ΔS [188].

In order to protect the La (Fe, Co)$_{13-x}$Si$_x$-based regenerators, tests of corrosion have been also performed by Balli et al [126] using different heat exchange fluids such as silicon oil, Zitrec (multiple usages) and water. Aiming to approach the operating conditions of functional devices, the experiments were performed in open atmosphere while the considered fluids are maintained in motion. The obtained results show that Zitric and particularly water alter drastically the LaFeCoSi matrix phase leading to marked mass losses. In contrast, the addition of only 3% of the anti-oxidant Noxal enables to completely protect the tested materials and reduces the mass loss to zero even after a long time immersion (see Fig.16-b). Additionally, the specific heat of the Noxal solution remains practically similar to that of water. It was also found that the silicon oil reduces significantly the corrosion effect. However, its specific heat that is only about 38 % of that shown by water (4.2 J/g K) [126] could drastically limit heat exchanges in the AMR-devices. In a following paper, a more detailed study regarding the corrosion behaviour of LaFeCoSi materials and Gd metal has been reported in Forchelet et al [190] using two distinct experiments that consist on immersions at both room temperature and 88 °C (accelerated test) during 336 h. The experimental tests were realized using several fluids including demineralized water, water (+Noxal 3 %), water (+Aquaris K-20 1%), water (+Sentinel 100X 1%), water (+BWT-SH1004 1%), water (+Aquaris R66) and Zitric S. It was found that the use of water mixed with very small amounts of some inhibitors such as Noxal, Sentinel X100 and Aquaris K-20 could be efficient to prevent mass losses induced by corrosion effects. On the other hand, Forchelet et al [190] have also studied the possibility to protect magnetocaloric materials through surface treatment or passivation by using oxalic acid solutions. For this purpose, Gd plates were immersed in different solutions of oxalic acid in deionized water showing different pH values up to 35 days [190]. The performed corrosion tests on a passivated Gd plate using the more aggressive demineralized water unveils that the passivation reduces drastically the mass losses caused by corrosion. In addition, the oxalic acid solution with pH = 0.75 seems to be more efficient and appropriate for passivation treatments. However, the protective oxide layer could significantly limits the heat exchanges between the refrigerant and the carrier fluid in the magnetic refrigerators. This is why additives-based heat transfer fluids are may be the best solution to prevent oxidation.

More recently, new studies in relation with the corrosion behaviour of La(Fe, Co)$_{13-x}$Si$_x$ based compounds were carried out [191-193]. In the work by Hu et al [191], it was found that the corrosion resistance of LaFe$_{13-x}$Si$_x$ compounds could be improved by introducing new elements in their matrix such as cobalt (Fe substitution) and carbon. All the corrosion experiments were performed in distilled water at room temperature using samples with only 1 working surface of 1 cm$^2$. The obtained results show that the combination of both Co and C in LaFe$_{13-x}$Si$_x$ drastically reduces the corrosion effect. For example, after 48 h immersion in distilled water, the corrosion rate of LaFe$_{10.87}$Co$_{0.63}$Si$_{1.5}$C$_{0.2}$ was 53.9 % lower than the mother compound LaFe$_{11.5}$Si$_{1.5}$. More interestingly, with the LaFe$_{10.87}$Co$_{0.63}$Si$_{1.5}$C$_{0.2}$ carbide, the mass loss is 33.3 % lower than the compound only with cobalt (LaFe$_{10.87}$Co$_{0.63}$Si$_{1.5}$). The high corrosion resistance of LaFe$_{10.87}$Co$_{0.63}$Si$_{1.5}$C$_{0.2}$ was confirmed by metallographs performed after immersion tests [191]. In a following paper, Hu et al [192] have studied the contribution of α-Fe and La-rich phases to the corrosion behaviour of LaFe$_{11.3}$Co$_{0.4}$Si$_{1.3}$C$_{0.15}$ and their effect on the magnetocaloric properties by using different tools such as scanning electron microscopy and magnetization measurements. It was found that the decrease of α-Fe and La-rich phase impurities which could be achieved through heat treatment improves markedly the corrosion resistance. In fact, α-Fe acts as the cathode while La-rich and matrix phases act as the anode to be corroded [192]. Furthermore, with increasing the annealing time, the amount of cathode decreases limiting the corrosion process in LaFe$_{11.3}$Co$_{0.4}$Si$_{1.3}$C$_{0.15}$. However, the corrosion resistance could be weakened if the La-rich phase is drastically reduced [192]. On the other hand, the corrosion resistance enhancement by reducing impurities also prevents a dramatic decrease of the entropy change. As reported in Ref.192, after 15 days immersion in distilled water, the maximum entropy change was reduced by 50 % for the sample annealed at 1353 K for 3 h. In contrast, the entropy change decreased only by about 16 % in the case of the sample annealed at 1353 for 7 days [192].

Recently, it was shown by Fujieda et al [193] that the corrosion resistance of LaFe$_{13-x}$Si$_x$ compounds could be significantly improved by reducing the dissolved oxygen (DO) concentration in the aqueous solutions that are used as heat exchange fluids. As reported in Ref.193, the aqueous corrosion of LaFe$_{13-x}$Si$_x$ was markedly reduced by decreasing the DO content. Additionally, the entropy change of LaFe$_{13-x}$Si$_x$ keeps high levels after immersion in deaerated distilled water with very low concentrations of DO. On the other hand, as pointed out by the authors [193], the immersion of LaFe$_{13-x}$Si$_x$ samples in distilled water increases their Curie temperature, which was attributed to the hydrogen absorption.

## V. ON THE ROTATING MAGNETOCALORIC EFFECT

In all above discussed materials, the magnetocaloric effect is obtained by subjecting the considered magnetic substance to a variable external magnetic field. However, in some magnetic materials that exhibit a large magnetocrystalline anisotropy, thermal effects could be also induced by rotating their single crystal between the easy and hard-axes in a constant magnetic field, as explained in Figure 17-a. Consequently, the cooling process could be achieved without the need to change continuosely the magnitude of the external magnetic field. More recently, a new design for the liquefaction of the hydrogen and helium was proposed, based on the rotating magnetoclaoric effect found in $HoMn_2O_5$ single crystals [196]. It is worth noting that this rotating (or anisotropic) magnetocaloric effect (RMCE) have attracted a little interest when compared with the conventional one [196-212]. This was mainly attributed to the fact that the contribution of the magnetocrystalline anisotropy to the MCE at the magnetic phase transition is much lower than that generated by the change in the magnetic order [197]. However, for different reasons, the implementation of the rotating MCE could revolutionize research and development on magnetic cooling technology for both low and room temperature applications: 1) In magnetic cooling systems using conventional MCE, the magnetization-demagnetization process generally requires a large mechanical energy for moving the active material in and out of the magnetic field zone, decreasing consequently the system efficiency. Hence, the use of the RMCE would enable the reduction of the energy absorbed by the cooling machine. 2) The implementation of such effect allows the conception of rotary magnetic refrigerators working at high frequency leading to a large cooling power. 3) The continuous variation of the magnetic field in cooling systems leads to the appearance of electric currents in metallic refrigerant materials. RMCE in a constant magnetic field eliminates the energy losses and additional works caused by the resulting eddy currents [158]. 4) It is known that rotary magnetic refrigerators are more efficient than reciprocating devices [48]. However, for rotary systems using the "standard MCE", the need to create a magnetic field gradient makes the design of the magnetic field source and consequently the cooling machine very complex. Therefore, the design of the machine can be drastically simplified by the implementation of materials exhibiting a large anisotropic MCE, since this kind of devices requires a simple constant magnetic field source (Fig.17-c) that would lead to more compact setups [196]. 5) The implementation of the RMCE can also be of benefit from an economical point of view, since the rotating motion can be easily realized with the help of cheaper circular motors.

The RMCE in terms of the entropy change ($\Delta S_R$) can be also determined from magnetization isotherms by using the Maxwell relation (Eq.9, section II-B). In this case, the rotating entropy change associated with the rotation motion from the hard axis (h) to the easy axis (e) in the field H can be expressed as follow:

$$\Delta S_{R,he} = \Delta S (H//e) - \Delta S (H//h) \qquad (40)$$

where the magnetic field is initially oriented along the hard-axis. $\Delta S (H//e)$ and $\Delta S (H//h)$ are the entropy changes when the magnetic field is applied along the easy and hard-directions, respectively. The rotating adiabatic temperature change $\Delta T_{ad,he}$ can be determined from the full entropies along the easy and hard-axes as demonstrated in Figure 17-b. In this case, $\Delta T_{ad,he}$ is given by:

$$\Delta T_{R,ad}(T,H) = [T(S)_{H//e} - T(S)_{H//h}]_S \qquad (41)$$

where S (H//e) and S(H//h) curves can be constructed from specific heat data with the help of equation 11 (section II-B).

More recently, several materials with large RMCE such as $RMnO_3$ and $RMn_2O_5$ multiferroics were mainly reported for cryogenic applications [210]. In addition to a conventional MCE that can be obtained by magnetizing these compounds along their easy-axes, meaningful RMCEs can be also generated by spinning them around the intermediate-axis in constant magnetic fields on account of their large magnetocrystalline anisotropy [210]. Particularly, the orthorhombic phases of $RMnO_3$ manganites unveil a large RMCE around the ordering point of $R^{3+}$ magnetic moments that is closer to 10 K. For example, the rotation of the orthorhombic $DyMnO_3$ single crystal in a constant magnetic field of 7 T within the bc-plane enables a maximum entropy change of 16.3 J/kg K and a maximum adiabatic temperature change of 11 K to be generated [207, 210]. In contrast, relatively low magnetic fields are required to achieve a large RMCE in $TbMn_2O_5$ single crystals [206]. In a constant magnetic field of 2 T, the adiabatic temperature change resulting from the rotation of $TbMn_2O_5$ crystals within the ac-plane reaches a maximum value of about 8 K, being much larger than that reported in $RMnO_3$ and other $RMn_2O_5$ oxides [210]. To learn more about the RMCE in $RMnO_3$ and $RMn_2O_5$ compounds, the interested reader is referred to the very recently reported review in Ref. 210.

In the room-temperature range, the RMCE has been reported in a limited number of single crystals. In the work by Nikitin et al [197], a giant RMCE was pointed out in a $NdCo_5$ single crystal. The latter unveils two spin reorientation transitions at $T_{SR1}$ = 250 K and $T_{SR2}$ = 290 K, leading to a large anisotropy of the magnetocaloric effect. Under a magnetic field of 1.3 T, the adiabatic temperature change resulting from the rotation of the $NdCo_5$ crystal between the a and c-axes reaches a maximum value of 1.6 K at 280 K. This large RMCE would enable to open the way for the implementation of $NdCo_5$ crystals in new types of room-temperature magnetic cooling systems. However, the difficulty associated with the crystals growth remains a serious obstacle to their utilization. Aiming to overcome this drawback, new alternatives such as textured polycrystalline materials have been suggested [211, 212]. In the work by Hu et al [211], the powder of $NdCo_4Al$ which presents spin reorientation temperatures at $T_{SR1}$ = 295 K and $T_{SR2}$ = 305 K, was oriented along the c-axis under a magnetic field of 0.5 T at 350 K by using the epoxy resign. Under a low magnetic field of 1 T, the textured $NdCo_4Al$ powder enables to generate a meaningful rotating entropy change of 1.3 J/kg K at 295 K. This is of great interest from both economical and practical points of view since the magnetic field-aligned technology is easy and low cost when compared with the preparation techniques for single crystals. Zhang et al [212] have proposed the textured DyNiSi polycrystalline material for low temperature RMCE-based magnetic refrigeration (around 10 K). Under a magnetic field of 5 T, the textured DyNiSi presents a maximum rotating entropy change of 17.6 J/kg K at 13 K. The associated adiabatic temperature change was found to be 10.5 K. The large RMCE makes the proposed refrigerant very promising for both cryogenic MCE-based devices and could be useful in some specific applications such as the liquefaction of helium and hydrogen (for example).

## VI. MULTILAYERED MAGNETOCALORIC REFRIGERANTS

Even though several magnetic materials showing giant magnetocaloric effects were reported, their working temperature range usually remains limited around the phase transition region. However, most of magnetic cooling systems utilize the AMR thermodynamic cycle to achieve large performance [52]. For this purpose, the used refrigerant must present excellent magnetocaloric properties over a wide temperature range. On the other hand, in an ideal Ericsson cooling cycle, the isothermal entropy change must remain unchanged over the considered working temperature range, as shown in Figure 18-a. Hence, an efficient refrigeration process in both AMR and Ericsson cycles cannot be performed only within a single magnetocaloric material. These constraints can be usually avoided by using composite refrigerants where several performant magnetocaloric materials are combined in order to build a multilayer regenerator efficiently working in the temperature range limited by their phase transition points [29-30, 67-68, 213]. Such refrigerants were proposed in the past for low temperature applications [213]. Hashimoto et al [213] have reported a sintered layer composed of $ErAl_{2.5}$, $HoAl_{2.5}$ and $Ho_{0.5}Dy_{0.5}Al_{2.5}$ with $T_C$ = 11, 26 and 33 K and mass ratios of 31.2, 19.8 and 49 %, respectively. The designed multilayer enables to cover the temperature range comprised between 10 and 40 K [213]. Later, several composites based on $R_{1-x}R'_x$ rare earths and other giant magnetocaloric materials such as $LaFe_{13-x}Si_x$ were proposed in the literature [29-30, 67-68].

As mentioned above, in an Ericsson cycle the isothermal entropy change must remain constant over the required temperature range. In order to meet this requirement, the constituent elements of the considered composite must be combined in optimum mass ratios where the accurate values can be obtained with the help of a specific numerical method [29-30, 67-68]. In this case, the isothermal entropy change $\Delta S_{Com}$ of a composite constituted of $n$ magnetocaloric materials in the $y_1, y_2...y_n$ proportions with Curie temperatures $T_C^1, T_C^2 ... T_C^n$ covering the suitable temperature range can be expressed as follow:

$$\Delta S_{Com} = \sum_{i=1}^{n} y_i \Delta S_i. \qquad (42)$$

Taking into account the fact that $\Delta S_{com}$ is constant over the working temperature range, equation 42 can be written as:

$$\sum_{j=1}^{n} y_j \left[ \Delta S_j (T_c^{i+1}) - \Delta S_j (T_c^i) \right] = 0, \text{ for } i = 1, 2, ..., n-1 \qquad (43)$$

where $\Delta S_j$ corresponds to the isothermal entropy change of the $j^{th}$ constituent. Considering the fact that $\sum_{j=1}^{n} y_j = 1$, the optimum mass ratios $y_1, y_2...y_n$ of each constituent can then be obtained by resolving the following matrix:

$$\begin{bmatrix} A11 & A12 & .... & .... & A1n \\ A21 & A22 & .... & .... & A2n \\ .... & .... & Aij & .... & Ain \\ An-11 & .... & .... & An-1n-1 & An-1n \\ 1 & .... & .... & 1 & 1 \end{bmatrix} \begin{bmatrix} y1 \\ y2 \\ yi \\ yn-1 \\ yn \end{bmatrix} = \begin{bmatrix} 0 \\ 0 \\ 0 \\ 0 \\ 1 \end{bmatrix} \qquad (44)$$

with $Aij = \Delta S_j (T_c^{i+1}) - \Delta S_j (T_c^i)$.

It is worth noting that the optimum mass ratios vary with the applied magnetic field. Hence, the multilayer's composition must be determined using the magnetic field of the magnetic cooling device [68]. For example, we report in Figure 18-b, the isothermal entropy change ($\Delta S_{Com}$) of a composite refrigerant based on $Gd_{1-x}Tb_x$ alloys that is built based on the above method and proposed in Ref.68. As shown, $\Delta S_{Com}$ remains practically constant in the temperature range close to room temperature (260-300 K).

## VII. CONCLUSIONS

Since the discovery of giant magnetocaloric effect in $Gd_5(Ge_{1-x}Si_x)_4$ compounds in the late of 1990s, a considerable effort was dedicated by worldwide research groups with the aim to provide more cheaper and efficient magnetocaloric materials for magnetic refrigeration applications. Currently, three families of magnetic materials including $R_{1-x}A_xMnO_3$ manganites, $La(Fe, Mn, Co, Si)_{13-x}Si_xH_y$ and $MnFe(P, As, Si, Ge)$ compounds were clearly identified to be promising alternatives for Gd-based alloys. Particularly, outstanding performances were recently reached by using $LaFe_{13-x}Si_xH_y$ hydrides as magnetic refrigerants, unveiling the bright future of magnetic cooling technology. Additionally, the direct implementation of $LaFe_{13-x}Si_x$ and $MnFeP_{1-x}As_x$ based compounds shows a constant increase in terms of thermodynamic performance, rendering the magnetic cooling closer to the commercialization phase. What remains now is to test these materials over a long period of time.

Regarding corrosion and mechanical brittleness issues, several research works are in progress and first encouraging results were obtained. However, the "magic" magnetocaloric material that exhibits a giant MCE over a wide temperature range (giant refrigerant capacity) combined with strong chemical and mechanical stabilities, low hysteresis, high thermal conductivity, high electrical resistance and low price has not yet been reported, opening the way for further investigations. In addition, although a big progress was made in going from the search for appropriate magnetocaloric materials to the design of efficient magnetic cooling devices, there are also still some technical issues to overcome such as the reduction of devices' weight (and size), providing systems with reasonable costs and the meet of industrial standards. In this context, the implementation of new materials presenting excellent magnetocaloric properties under low magnetic fields would enable to markedly reduce the quantity of permanent magnets used by field sources in the magnetocaloric devices. This will positively impact the cost as well as the size (and weight) of magnetic refrigerators.

**ACKNOWLEDGMENTS**

We acknowledge the financial support from NSERC (Canada), FQRNT (Québec), CFI, CIFAR, Canada First Research Excellence Fund (Apogée Canada), and the Université de Sherbrooke.

M. Balli would like to thank the Grenoble Institute of Technology (France) and especially the director of G2Elab, Prof. James Roudet, for having hosted him as Invited Scientist during the year 2016.

**FIGURE CAPTIONS**

**FIG.1.** A view of the magnetic cooling system designed by the University of Applied Sciences of Western Switzerland (Balli et al). The magnetocaloric effect is induced by a permanent magnets-based field source. The magnetization-demagnetization of magnetocaloric material (MCM) is performed by a linear actuator[8, 49].

**FIG.2.** Full entropy of Gd as a function of temperature under 0 and 5 T, deduced from MFT theory (see section II-C). As shown, the change in the magnetic order under the application of an external magnetic field gives rise to the magnetocaloric effect phenomenon. For Gd, $g_J = 2$, $J = 7/2$ and $T_C = 294$ K [52, 53].

**FIG.3.** Experimental [65] (triangles) and calculated (solid line) magnetizations of $La_2NiMnO_6$ double perovskite as a function of temperature under 5 T. In the calculation, the Lande factor $g_J$ is assumed to be 2, $J = 2.75$ and $T_C = 280$ K.

**FIG.4.** Temperature dependence of magnetization in MnAs under 0 and 5 T, obtained from Bean-Rodbell Model [53, 63, 69-76] for $T_0 = 285$ K, $J = 3/2$, $g_J = 2.26$ and $n = 2.28$. As shown, $T_C$ increases with increasing magnetic field as a consequence of the magneto-structural interplay in MnAs.

**FIG.5.** Temperature dependence of the calculated full entropy at 0 and 5 T for MnAs, using Bean-Rodbell model [53, 63, 69-76]. Inset: deduced isothermal entropy change as a function of temperature under 5 T. The used parameters are $T_0 = 285$ K, $T_D = 310$ K, $J = 3/2$, $g_J = 2.26$ and $n = 2.28$.

**FIG.6:** Isothermal magnetization curves of a MnAs sample[90] around its $T_C$ = 317 K. Inset: deduced isothermal entropy change by directly integrating the Maxwell relation up to 7 T. As shown, the direct integration of Maxwell relation without taking into account the hysteresis effect yields to unreasonable values of $-\Delta S$ (more than 130 J/kg K under 7 T).

**FIG.7.** Resulting magnetic field (triangles) as a function of temperature inside a sample of Gd (2*2*2 mm$^3$)[90] subjected to an external magnetic field of 1 T (dashed line).

**FIG.8.** Enlarging the working temperature range of Gd using $Gd_{1-x}Tb_x$ alloys (data taken from Ref. 53).

**FIG.9.** Temperature dependence of the isothermal entropy change in $LaFe_{11.7}Si_{1.3}$ under 2 and 5 T (data taken from Ref. 31).

**FIG.10.** (a) Isothermal entropy change as a function of Curie temperature for $La(Fe, Co)_{13-x}Si_x$ [29] and $LaFe_{13-x}Si_xH_y$ [27] compounds under a magnetic field change of 2 T. (b) Adiabatic temperature change as a function of Curie temperature under 2 T for $LaFe_{13-x}Si_xH_y$ [27]. (c) Effective magnetocaloric effect (see section II-D) as a function of temperature for a sample of $La(Fe, Co)_{13-x}Si_x$ [126].

**FIG.11.** Example of a regenerator based on $La(Fe, Co)_{13-x}Si_x$ materials, co-designed by the University of Applied Sciences of Western Switzerland (HES-So) and Vacuumshmelze company. The regenerator is obtained by the powder metallurgy technique [32]. "Red zones" unveil the weak resistance of these materials against corrosion phenomena (see section IV).

**FIG.12.** Maximum obtained temperature difference (span) between hot and cold sources by using single or multilayer $La(Fe, Co)_{13-x}Si_x$ as refrigerants in magnetic cooling devices. A multilayer refrigerant combines several compounds with different Curie points $T_C$ (see section VI). More details are also given in table 2.

**FIG.13:** Transition temperatures and isothermal entropy changes (under 2 T) of (a) $Mn_{1.1}Fe_{0.9}P_{1-x}Ge_x$ [172] (b) $Mn_{1.2}Fe_{0.75-y}Ni_yP_{0.5}Si_{0.5}$ [170], (c) $Mn_{2-y}Fe_yP_{0.75}Ge_{0.25}$ [172] and (d) $Mn_{1.2}Fe_{0.8-z}Ru_zP_{0.5}Si_{0.5}$ [170] compounds.

**FIG.14:** Transition temperatures and isothermal entropy changes (under 2 T) of (a) $Mn_xFe_{1.95-x}P_{0.5}Si_{0.5}$ [165] (b) $Mn_{1.25}Fe_{0.7}P_{1-y}Si_y$ [165] compounds. (c) and (d) their adiabatic temperature change (under 1.1 T), respectively.

**FIG.15.** Obtained maximum temperature span using magnetocaloric oxides (singles or multilayers) as refrigerants in functional magnetic cooling devices. For more details, see table 3.

**FIG.16.** (a) Degradation of a La (Fe, Co)$_{13-x}$Si$_x$-based regenerator only a few days after its implementation in the magnetic cooling device described in Ref. 49. (b) Reduction of the La (Fe, Co)$_{13-x}$Si$_x$ corrosion by using additives such as Noxal [126].

**FIG.17.** (a) Generation of the magnetocaloric effect by rotating single crystals between their hard and easy axes. (b) Determination of the rotating adiabatic temperature and entropy changes from the full entropy following the hard and easy axes of a single crystal. (c) A design for the liquefaction of the helium and hydrogen by using the rotating MCE of HoMn$_2$O$_5$ [196].

**FIG.18.** (a) Principle of the Ericsson thermodynamic cycle. (b) The resulting entropy change of the composite Gd/Gd$_{0.7}$Tb$_{0.3}$/Gd$_{0.5}$Tb$_{0.5}$ as a function of temperature under 1 and 2 T (data taken from Refs. 53 and 68).

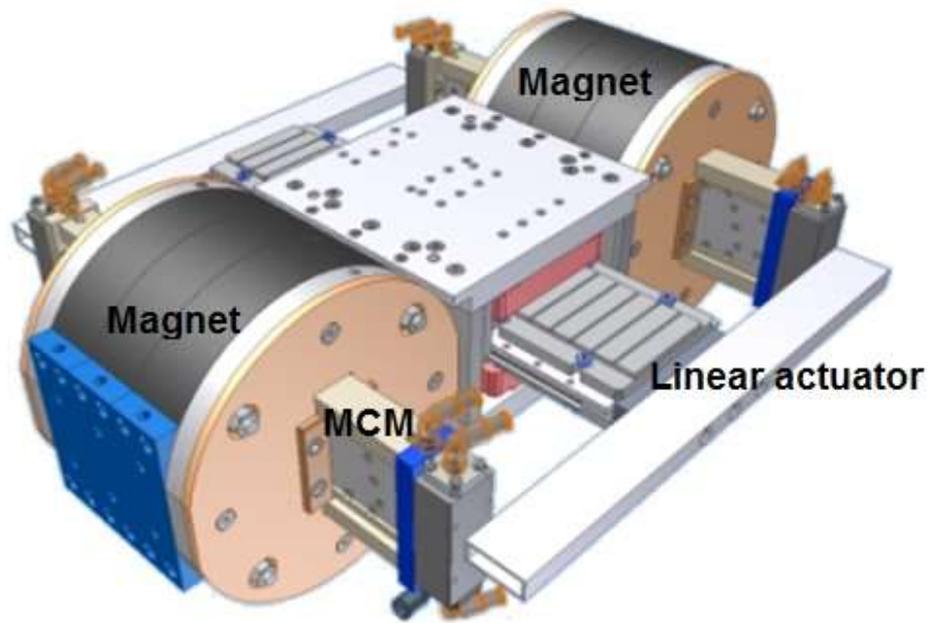

**Figure 1**

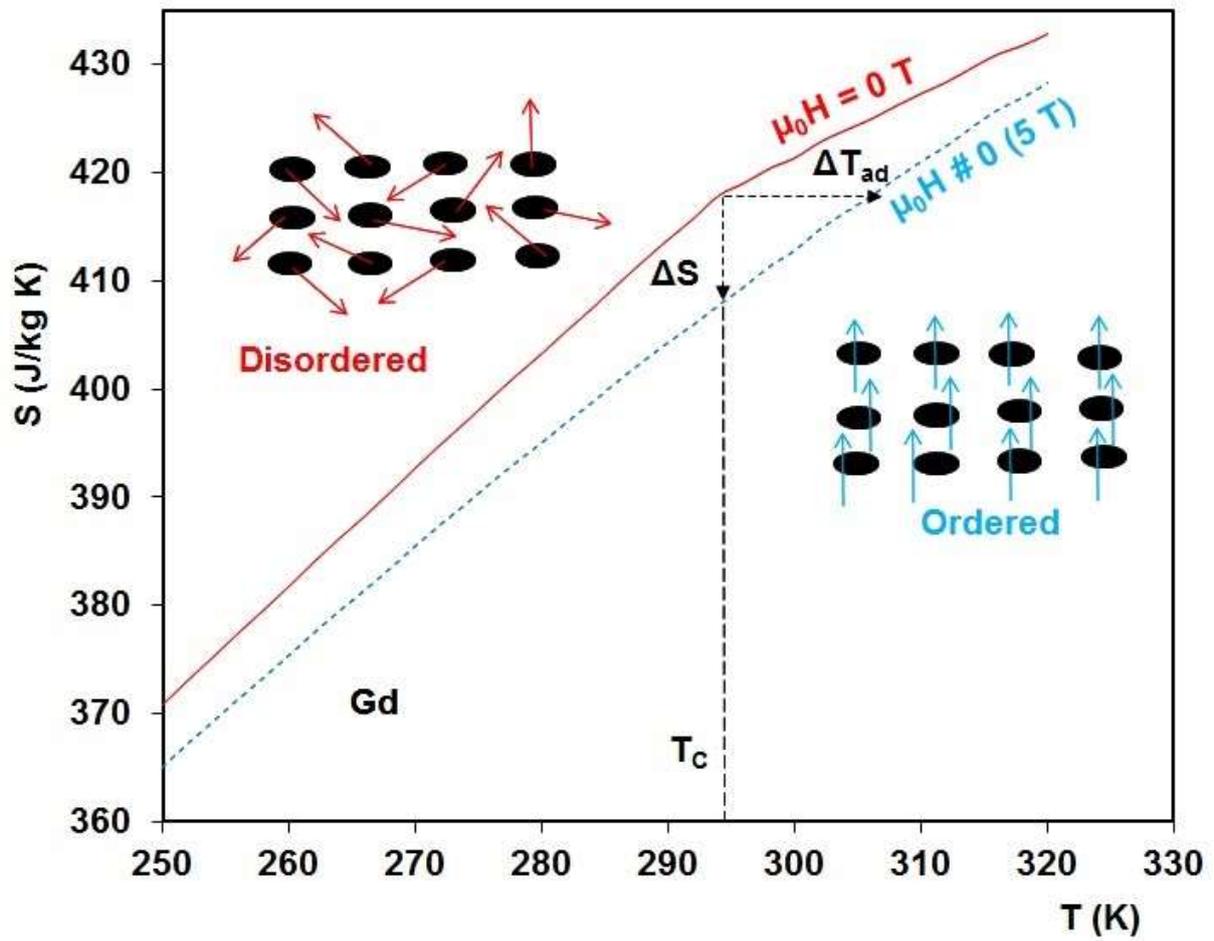

**Figure 2**

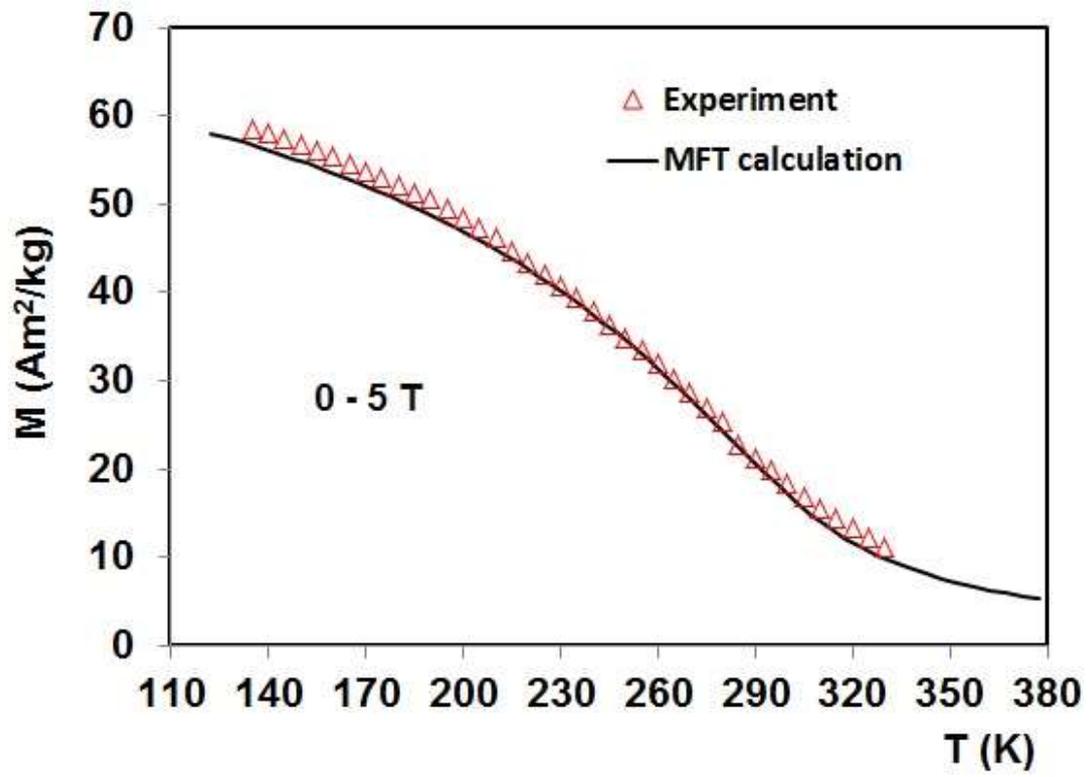

**Figure 3**

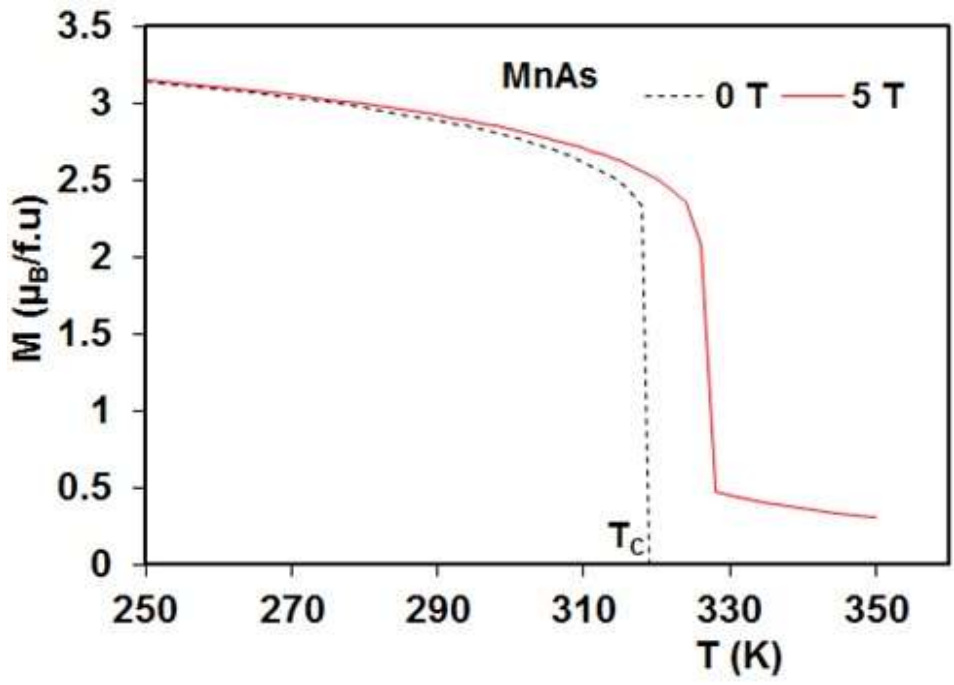

**Figure 4**

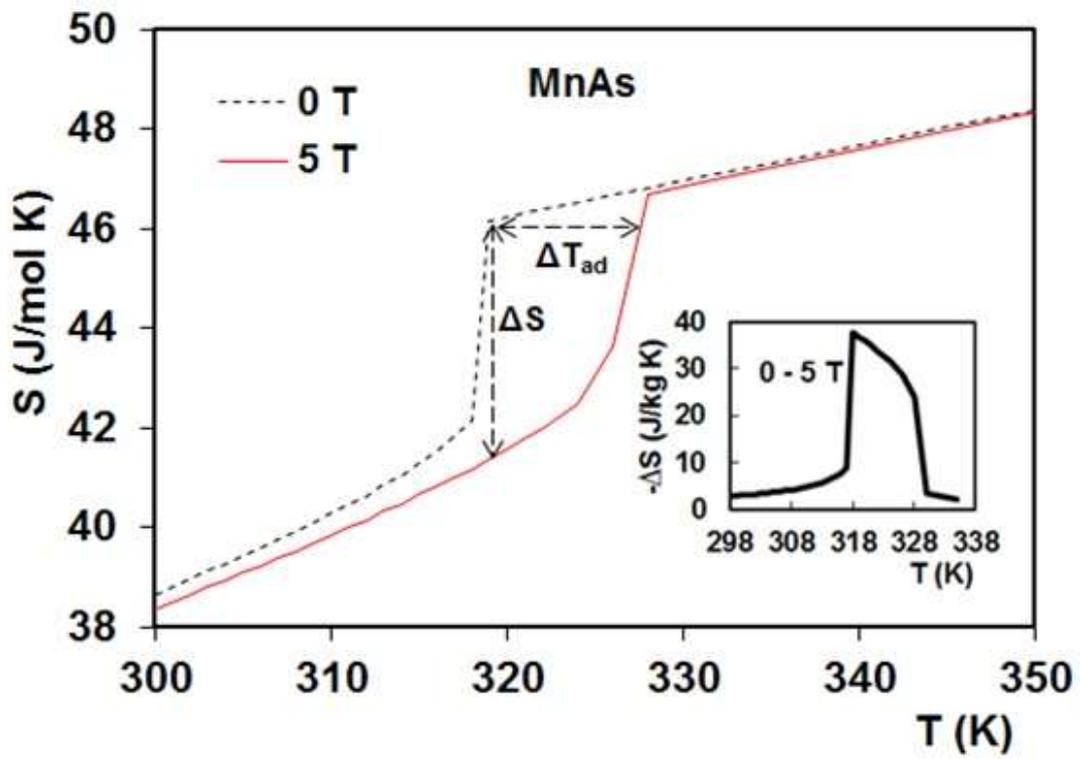

**Figure 5**

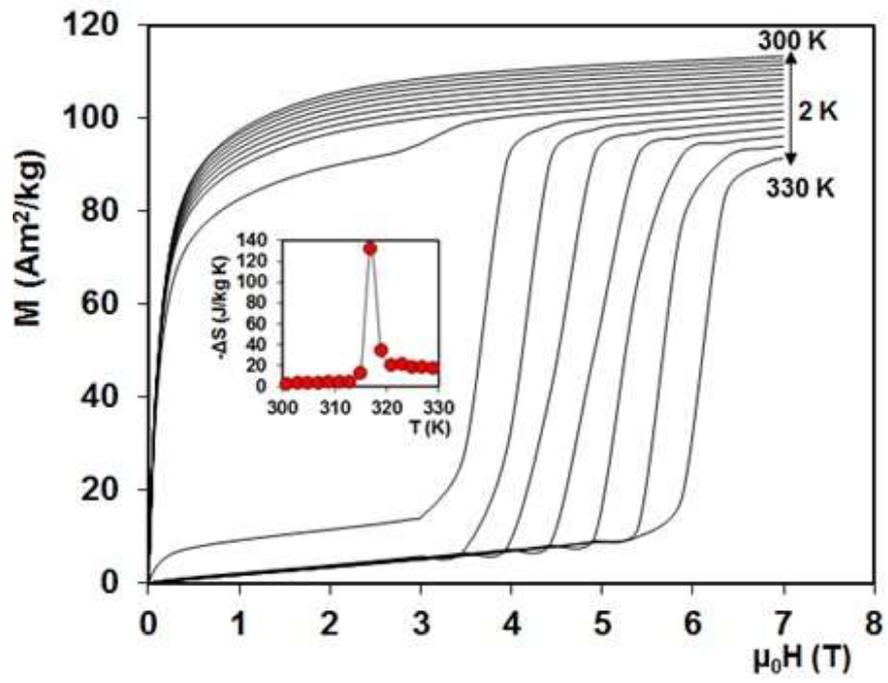

**Figure 6**

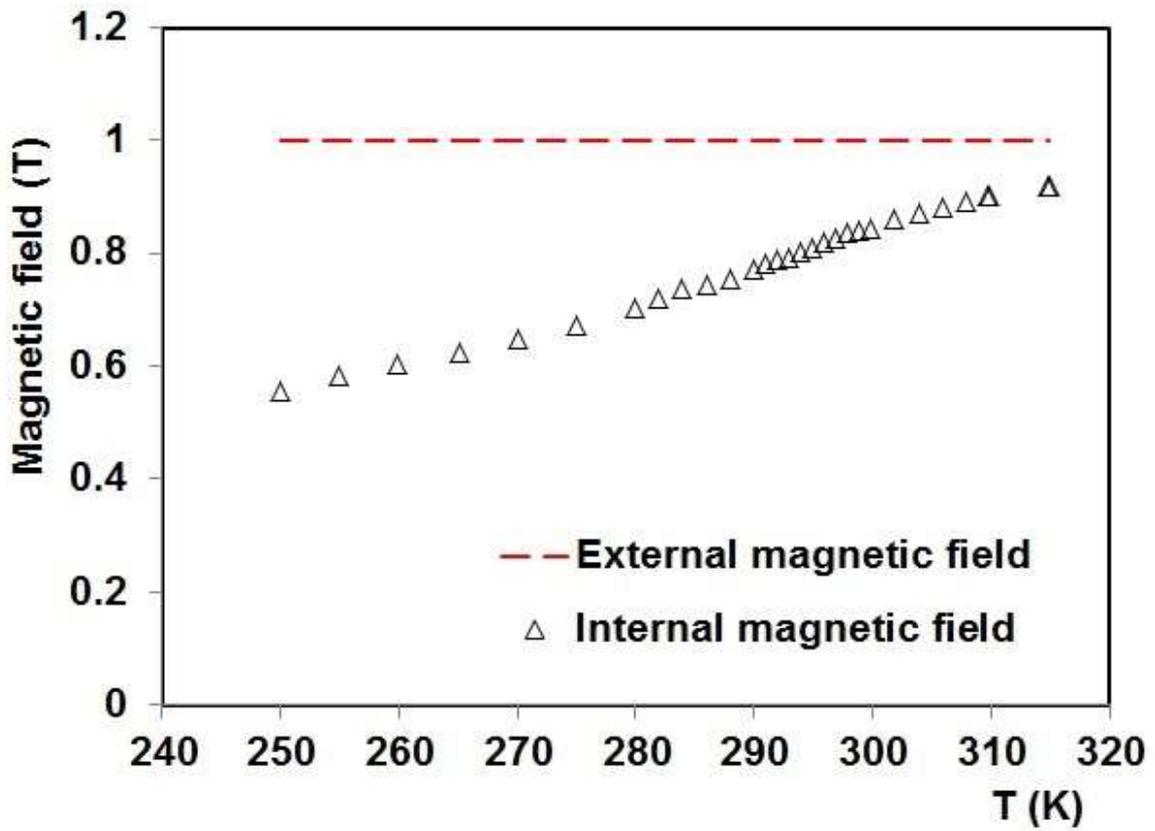

**Figure 7**

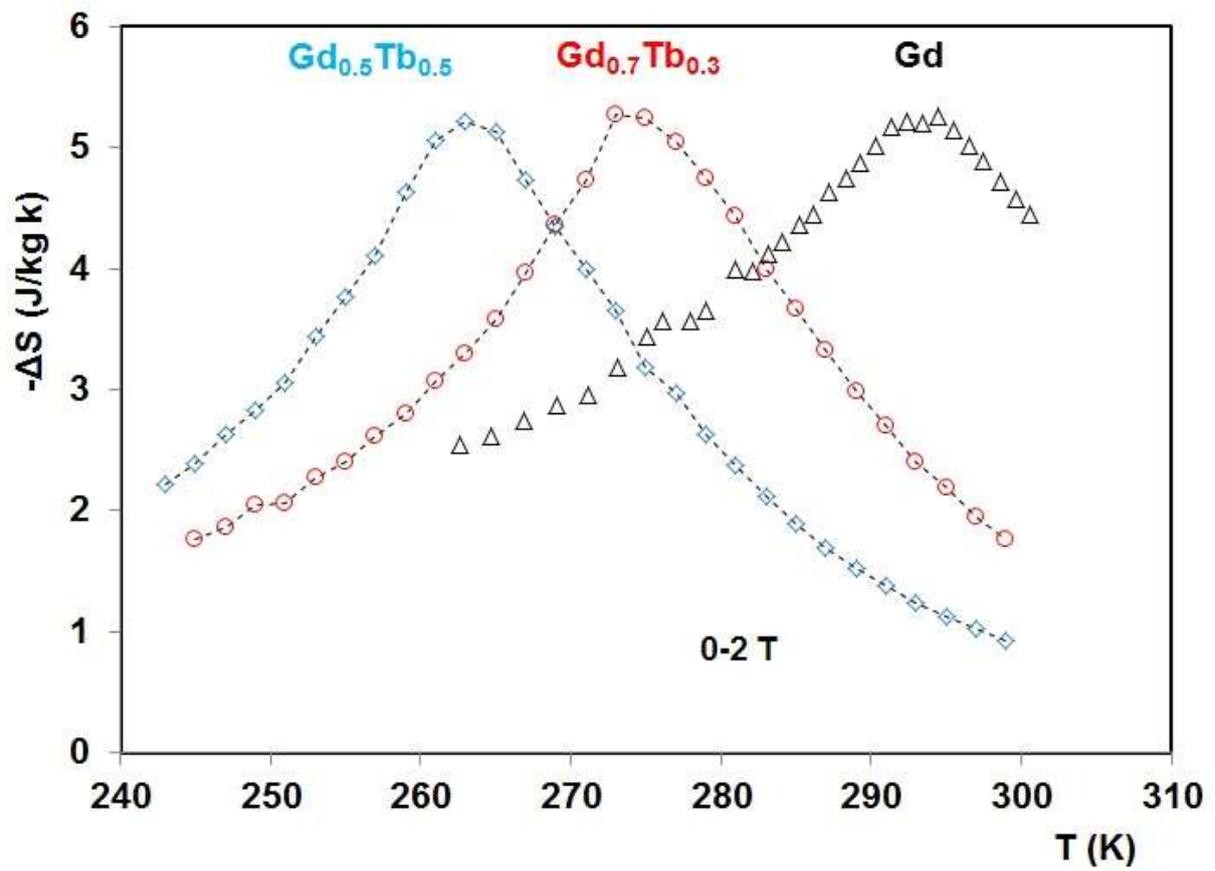

Figure 8

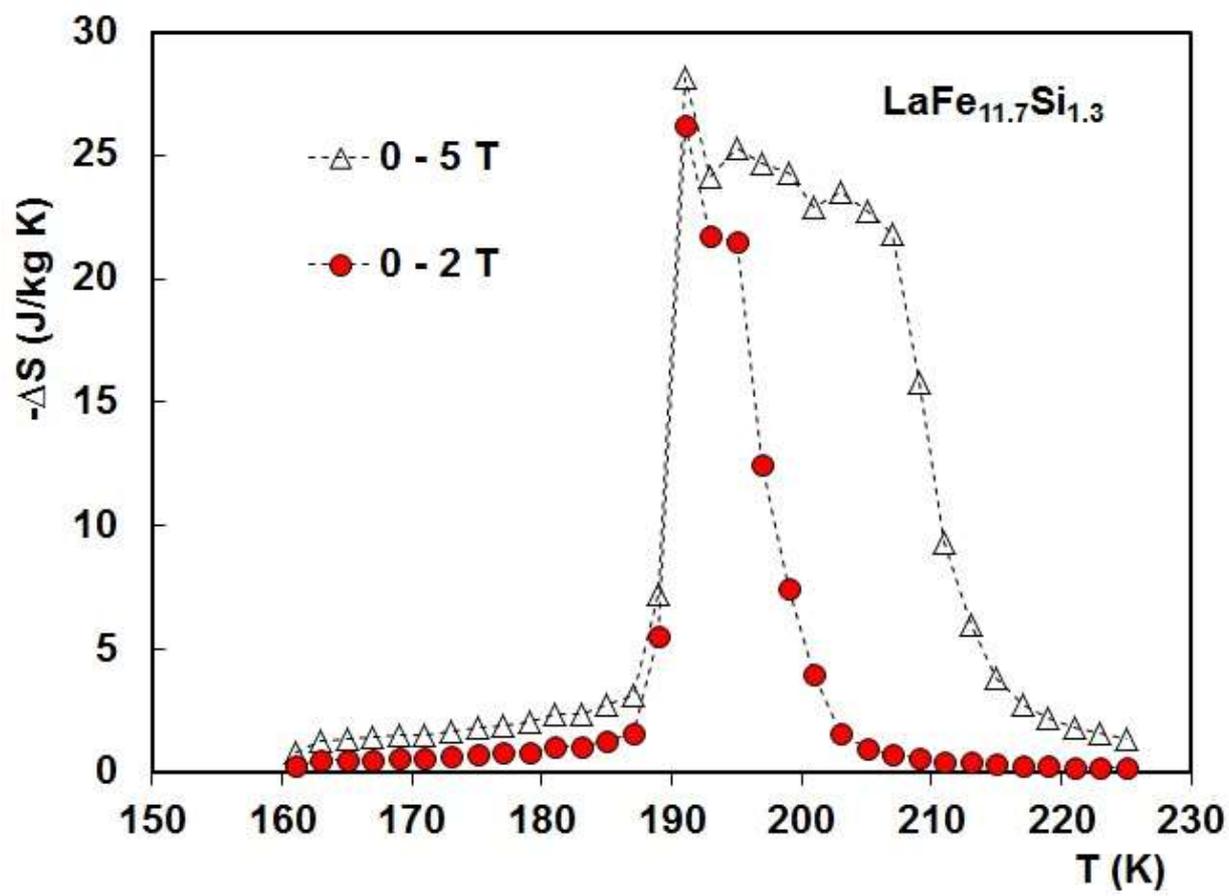

**Figure 9**

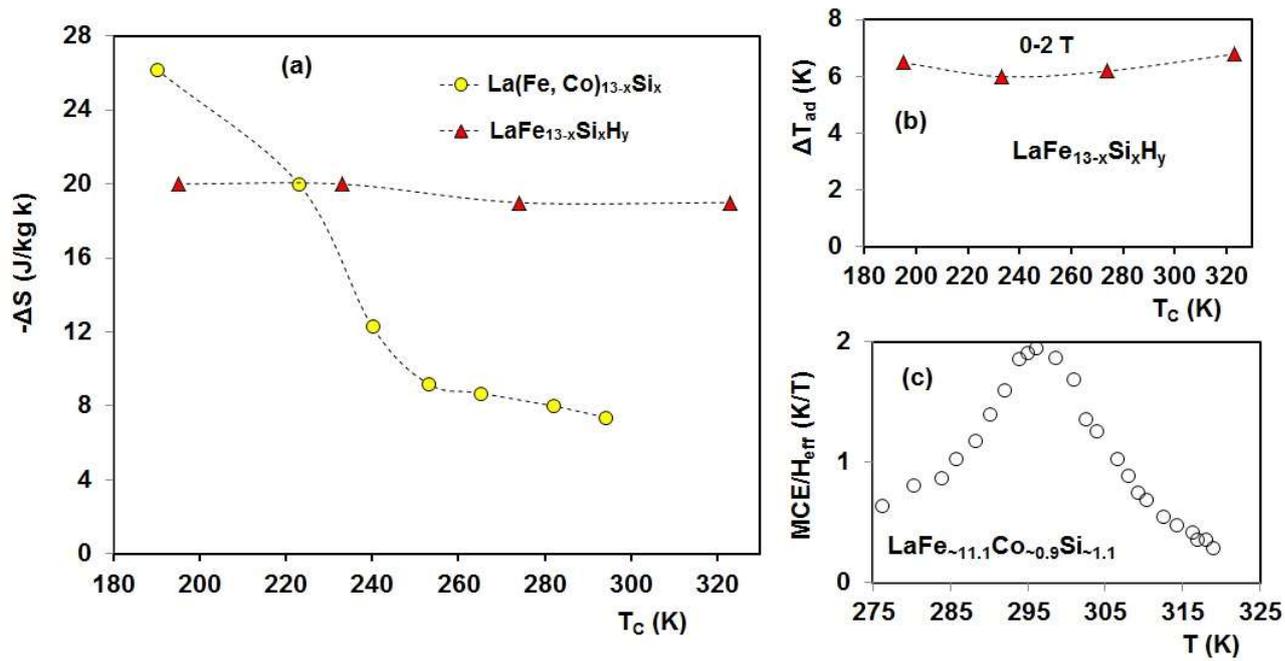

Figure 10

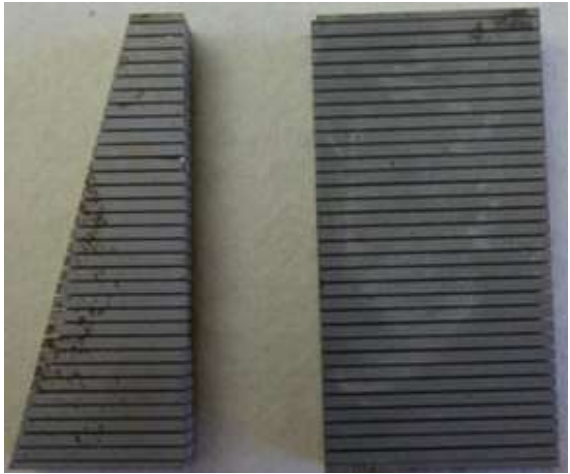

**Figure 11**

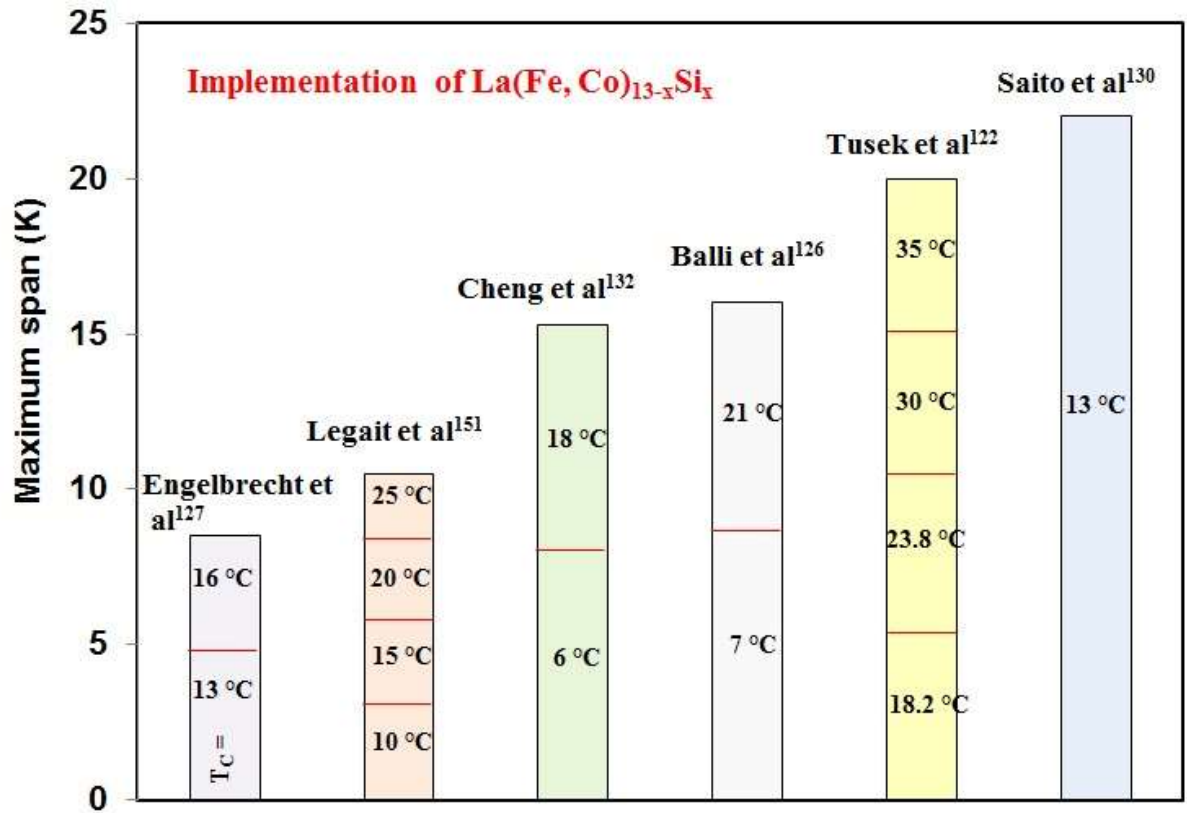

**Figure 12**

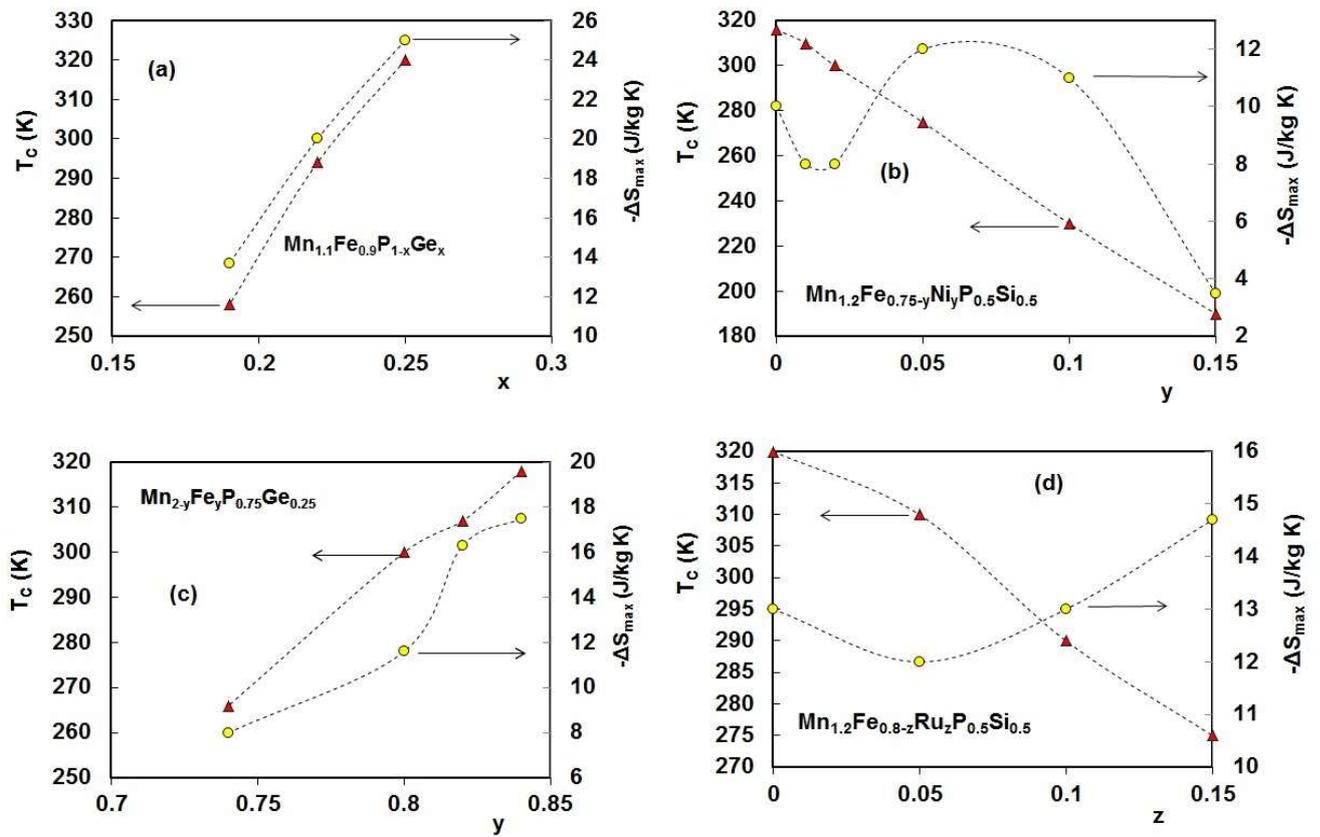

**Figure 13**

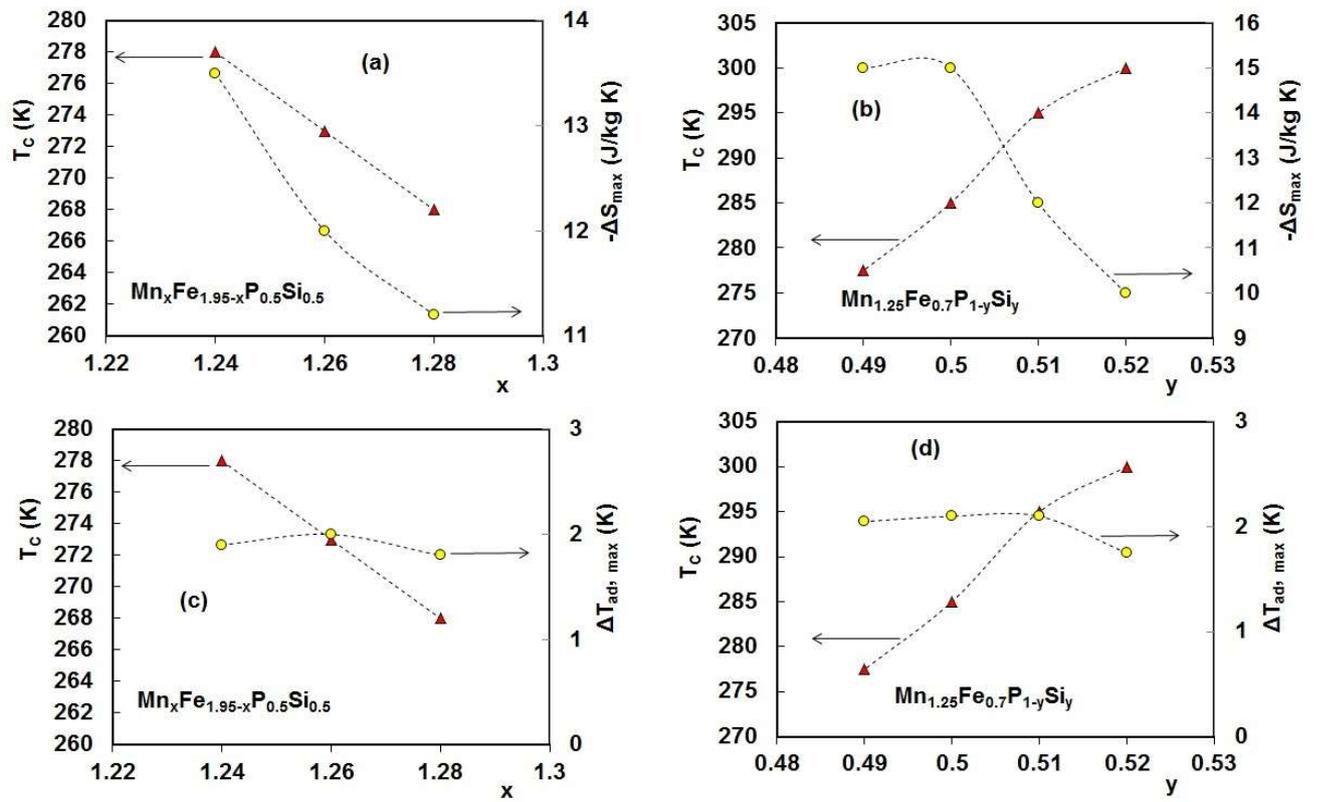

**Figure 14**

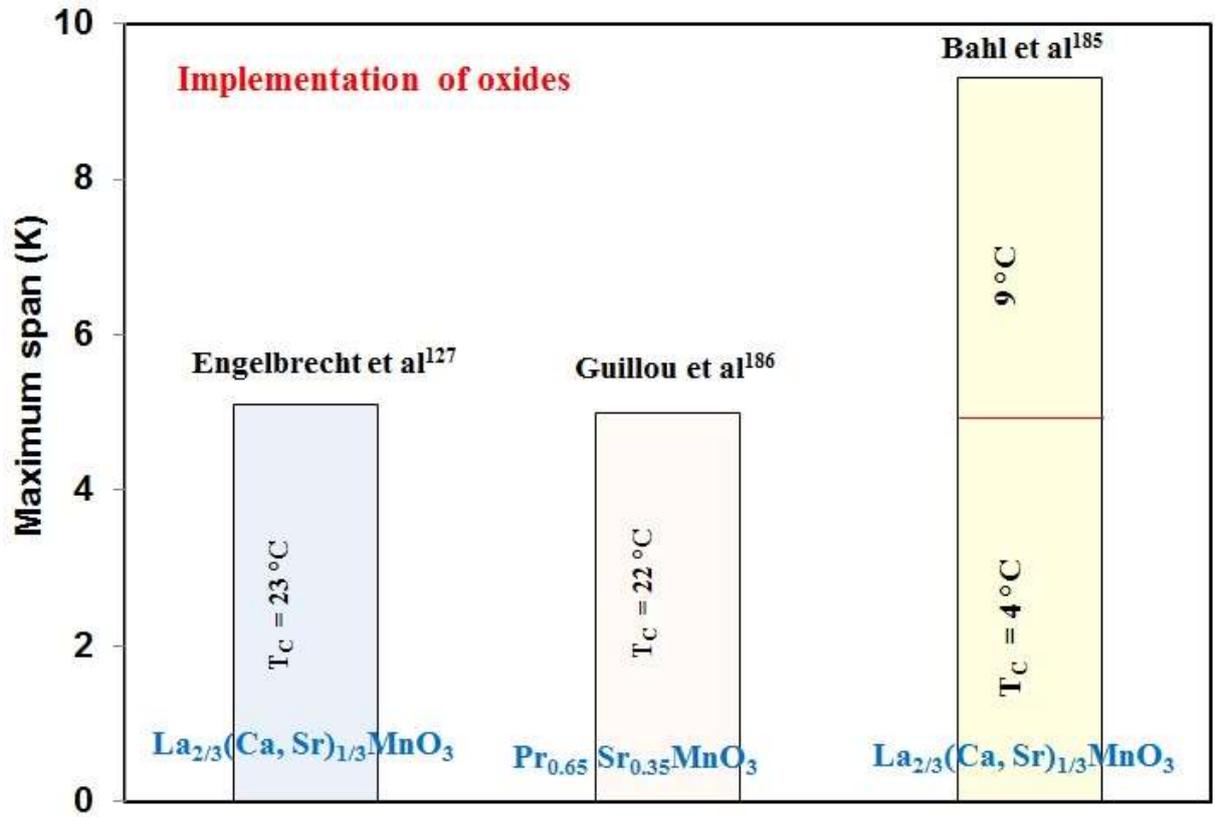

**Figure 15**

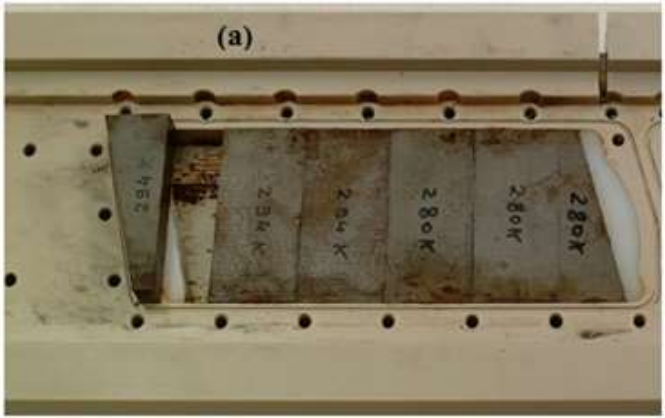 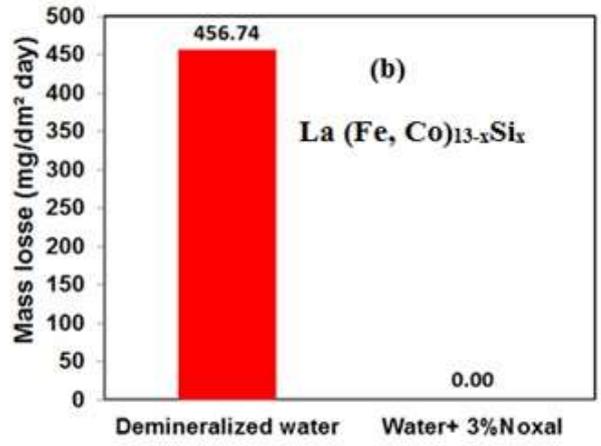

Figure 16

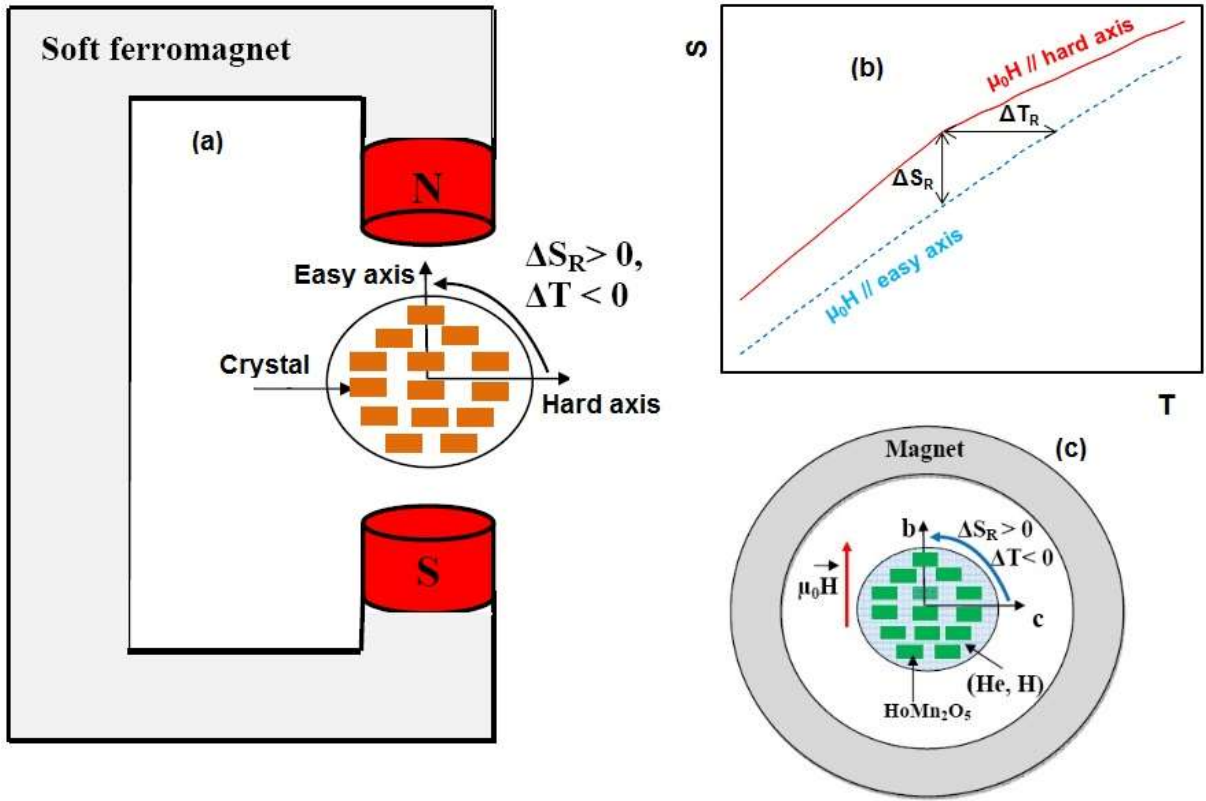

**Figure 17**

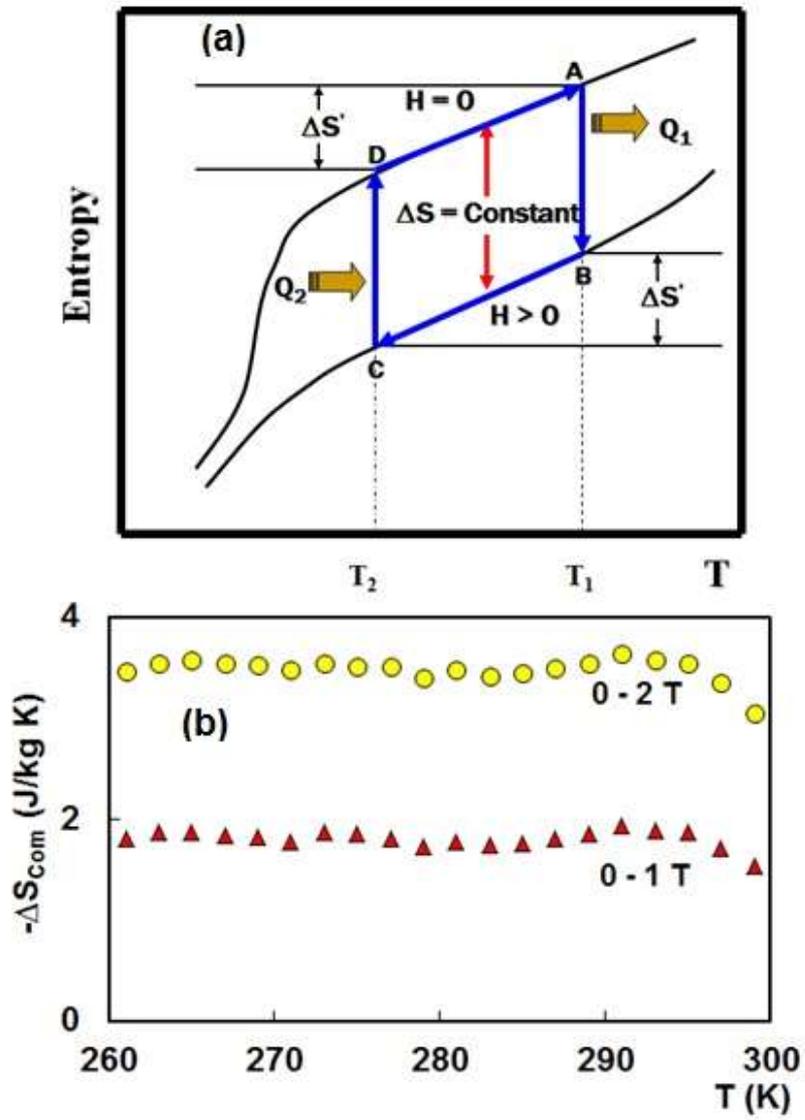

**Figure 18**